\documentclass[fleqn,usenatbib]{mnras}

\usepackage{newtxtext,newtxmath}
\usepackage[T1]{fontenc}
\usepackage{graphicx}	% Including figure files
\usepackage{amsmath}	% Advanced maths commands
\usepackage{multirow}
\usepackage{caption}
\usepackage{subcaption}
\usepackage{float}

\DeclareRobustCommand{\VAN}[3]{#2}
\let\VANthebibliography\thebibliography
\def\thebibliography{\DeclareRobustCommand{\VAN}[3]{##3}\VANthebibliography}

\title[Photometric classification of astronomical objects]{Machine learning-based photometric classification of galaxies, quasars, emission-line galaxies, and stars}

\author[Zeraatgari et al.]{
Fatemeh Zahra Zeraatgari,$^{1}$\thanks{E-mail: fzeraatgari@xjtu.edu.cn} 
Fatemeh Hafezianzade,$^{2}$
Yanxia Zhang,$^{3}$\thanks{E-mail: zyx@bao.ac.cn} 
Liquan Mei,$^{1}$\thanks{E-mail: lqmei@mail.xjtu.edu.cn} 
Ashraf Ayubinia,$^{4}$
\newauthor
Amin Mosallanezhad,$^{1}$
and Jingyi Zhang$^{3}$
\\
$^{1}$School of Mathematics and Statistics, Xi'an Jiaotong University, Xi'an, Shaanxi 710049, PR China\\
$^{2}$Department of physics, Institute for Advanced Studies in Basic Sciences, Zanjan, 45195-1159, Iran\\
$^{3}$CAS Key Laboratory of Optical Astronomy, National Astronomical
 	Observatories, Beijing, 100101, China\\
$^{4}$Department of Physics and Astronomy, Seoul National University, Seoul 08826, Republic of Korea}

\date{Accepted XXX. Received YYY; in original form ZZZ}

\pubyear{2023}

\begin{document}

%\label{firstpage}
\maketitle

\begin{abstract}
This paper explores the application of machine learning methods for classifying astronomical sources using photometric data, including normal and emission line galaxies (ELGs; starforming, starburst, AGN, broad line), quasars, and stars. We utilized samples from Sloan Digital Sky Survey (SDSS) Data Release 17 (DR17) and the ALLWISE catalog, which contain spectroscopically labeled sources from SDSS.
Our methodology comprises two parts. First, we conducted experiments, including three-class, four-class, and seven-class classifications, employing the Random Forest (RF) algorithm. This phase aimed to achieve optimal performance with balanced datasets. In the second part, we trained various machine learning methods, such as $k$-nearest neighbors (KNN), RF, XGBoost (XGB), voting, and artificial neural network (ANN), using all available data based on promising results from the first phase.
Our results highlight the effectiveness of combining optical and infrared features, yielding the best performance across all classifiers. Specifically, in the three-class experiment, RF and XGB algorithms achieved identical average F1 scores of 98.93 per~cent on both balanced and unbalanced datasets. In the seven-class experiment, our average F1 score was 73.57 per~cent. Using the XGB method in the four-class experiment, we achieved F1 scores of 87.9 per~cent for normal galaxies (NGs), 81.5 per~cent for ELGs, 99.1 per~cent for stars, and 98.5 per~cent for quasars (QSOs).
Unlike classical methods based on time-consuming spectroscopy, our experiments demonstrate the feasibility of using automated algorithms on carefully classified photometric data. With more data and ample training samples, detailed photometric classification becomes possible, aiding in the selection of follow-up observation candidates.
\end{abstract}


%%%%%%%%%%%%%%%%%%%%%%%%%%%%%%%%%%%%%%%%%%%%%%%%%%

%%%%%%%%%%%%%%%%% BODY OF PAPER %%%%%%%%%%%%%%%%%%

\section{Introduction}
When it comes to data in astronomy, it is typically collected from a diverse range of celestial objects. The classification of the data is a critical component of the astronomical data analysis process. By categorizing the data into different classes, we can gain valuable insights into the underlying physical processes and properties of the objects under investigation. As an example, in recent years there has been an influx of optical and near-infrared sky surveys (e.g. Sloan Digital Sky Survey (SDSS; \citealt{York2000}), Panoramic Survey Telescope and Rapid Response System (Pan-STARRS; \citealt{Chambers2016}), Dark Energy Survey (DES; \citealt{Reed2015, Reed2017, Reed2019, Yang2019}), Hyper Suprime-Cam (HSC) Subaru Strategic Program \citep{Aihara2018}, Two Micron All Sky Survey (2MASS; \citealt{Skrutskie2006}), UKIRT Infrared Deep Sky Survey (UKIDSS; \citealt{Lawrence2007}), VISTA Hemisphere Survey (VHS; \citealt{McMahon2012}), Visible and Infrared Survey Telescope for Astronomy (VISTA), Wide-field Infrared Survey Explorer (WISE; \citealt{Wright2010})). As a result, photometric data have gained a great deal of attention for the purposes of mapping and classifying astronomical sources. Thanks to large-area multiwavelength surveys, our knowledge and comprehension of the Universe and its diverse constituents have been significantly enhanced.

The ability to differentiate between point sources (stars and quasars) and extended sources (galaxies), is essential for astronomers in identifying various astronomical objects. Additionally, distinguishing normal galaxies from active or emission line galaxies is crucial for advancing our comprehension of various physical processes at play. One common approach to classify distinct sources from an image is by using morphology-based classification, which distinguishes between stars and galaxies (\citealt{Lopez-Sanjuan2019} and references therein), as well as passive and active galaxies (\citealt{Wilman2013, Man2021}). However, this approach has a disadvantage as it makes it impossible to distinguish between point sources like stars and quasars in a single image \citep{Fotopoulou2018}. Moreover, classifying active versus passive galaxies based entirely on their morphology can be challenging as some active galaxies can have a similar morphology to passive galaxies, and vice versa \citep{Tamburri2014}. Accurate classification often requires additional techniques such as spectroscopic analysis (\citealt{Baldwin1981, Veilleux1987, Kauffmann2003, Tanaka2012}), which detects ionized gas emitting strong emission lines from objects that have intense star formation or accreting black holes at their centers \citep{Conselice2003, Wen2014}. This method is effective but requires a significant investment of telescope time and is limited to small samples of objects. Additionally, spectroscopy alone cannot always distinguish between active and passive galaxies, as some passive galaxies can have low-level activity that cannot be detected.
Furthermore, some objects, including both passive and active galaxies, may be difficult to detect because they do not exhibit strong spectral features \citep{Zaritsky1995}.

Alternatively, astronomers often utilize photometric data, including colors and spectral energy distributions to differentiate between different sources. This approach offers a more efficient and cost-effective method of classifying galaxies, providing informative findings into their general properties, age, metallicity, and star formation history through analysis of colors and brightness \citep{Conti2003, Assef2010, Monachesi2012, Wang2022}.
Object selection, whether using WISE colors alone or in combination with other catalogues, primarily relies on utilizing color information to distinguish between active galactic nuclei (AGN)/quasars and stars or passive galaxies \citep{Richards2002, Richards2005, Schneider2007, Schneider2010, Jarrett2011, Jarrett2017, Stern2012, Mateos2012, Wu2012, Edelson2012, Goto2012, Assef2013, Yan2013, Tu2013, Chung2014, Nikutta2014, Ferraro2015, Secrest2015}.

In addition, wide-field surveys such as SDSS and WISE have played a crucial role in generating vast amounts of photometric information, enabling researchers to measure the properties of various astronomical objects. Machine learning-based algorithms have proven to be effective in handling and classifying this information. 
A majority of the literature on these two surveys in the case of photometric classification using machine learning, beginning from \citet{Suchkov2005}, have focused on broad classifications or specific subtypes of astronomical objects. For instance, some studies have examined separating stars from galaxies (\citealt{Ball2006, Vasconcellos2011, Kovacs2015}), distinguishing star/galaxy/QSO (\citealt{Krakowski2016}; \citealt{Kurcz2016}; \citealt{Nakoneczny2019, Nakoneczny2021}; \citealt{Clarke2020}; \citealt{Cunha2022}; \citealt{Chaini2022}), examining ELGs (AGN/non-AGN, Seyfert I/Sefert II) \citep{Cavuoti2014}, as well as classifying AGNs (X-ray AGNs, IRAGNs, radio-selected AGNs) \citep{Chang2021}. Despite significant progress in the field, a gap remains in the literature regarding the detailed classification of all astronomical objects using solely photometric data.
To address the current gap in the literature, we present a comprehensive and adaptable classification system that aims to provide a solution to this challenge, considering our four main objectives.

As our first objective, we propose to develop a reliable classification system that will accurately classify sources of all types, including subclasses of galaxies, and determine stars from quasars using only photometric data. By taking this approach, we will be able to separate point sources from extended sources, and distinguish active galaxies from passive galaxies.
Furthermore, by automating and facilitating source identification, 
our proposed classification system serves as a tool for analyzing large datasets, 
streamlining the identification process, and enhancing efficiency. This is mainly because with the advent of 
future photometric surveys, such as Large Synoptic Survey Telescope 
(LSST; \citealt{LSST2009, LSST2012, Ivezic2019}), a large amount of 
data will be generated, and analysis methods will need to be automated to uncover new scientific insights.

Secondly, our proposed approach has the potential to make significant contributions to this field by providing a fine-grained classification of astronomical objects using only photometric data. In this study, we specifically focus on the classification of stars, quasars (QSOs), emission-line galaxies (ELGs), and normal galaxies (NGs). These object categories were chosen based on their relevance to the research community and their distinct characteristics. Stars, as fundamental objects in astrophysics, play a crucial role in understanding various astrophysical processes. Quasars, known for their unique high-energy emission properties, provide significant information about the most energetic phenomena in the Universe. Emission-line galaxies, with their prominent emission lines, present significant clues about the underlying physical mechanisms at work. Additionally, the category of emission line galaxies encompasses a diverse range of object types that require comprehensive classification. Furthermore, in our study, we extend the classification to include additional subclasses, resulting in a total of seven distinct groups: normal galaxies (NGs), starforming galaxies (SF), starburst galaxies (SB), active galactic nuclei (AGN), broad line galaxies (BL), in addition to the aforementioned stars and quasars. By specifically addressing these seven groups, our proposed classification system bridges a significant void in the literature and offers a detailed and versatile approach for accurately classifying astronomical objects using photometric data. This comprehensive classification system will enhance our understanding of the properties and characteristics of these objects, enabling new scientific discoveries and advancements in the field.

Thirdly, in order to effectively classify the astronomical sources in our study, 
we decide to implement supervised machine learning techniques and a novel approach 
that incorporates photometric information from spectroscopically identified sources. 
By doing so, we can reliably classify all classes of astronomical sources with an acceptable degree of accuracy. 
This methodology has the potential to significantly advance our understanding of the properties 
and characteristics of these celestial objects, as well as contribute to the development of 
more efficient and effective methods for identifying and classifying astronomical sources based 
exclusively on photometric data. 
The final objective of this study is to investigate efficient selection of astronomical objects by 
engaging a variety of machine learning models, including $k$-nearest neighbors (KNN), random forest (RF), XGBoost (XGB), voting, and artificial neural network (ANN). 
This methodology enables us to enhance the accuracy of our analyses and attain 
a comprehensive understanding of the data for precise classification of objects.
In order to extract optical and IR photometric information, we cross-match SDSS and ALLWISE catalogues. 
To extract QSOs with the highest probability of reliability, Milliquas is utilized.
During the study, we use a two-step methodology to guarantee the precision of our classifiers, 
and we meticulously adhere to this procedure to maintain uniformity in our outcomes.

The structure of this paper can be summarized as follows:
Section \ref{sec:data} describes the sample used in this paper.
The adopted methods are briefly introduced in Section \ref{sec:method}.
Section \ref{sec:result} discusses the results of the classifiers in detail.
An overall comparison of the present work with other similar studies 
can be found in Section \ref{sec:discussion}.
The summary and conclusion of this study are provided in Section \ref{sec:conclusion}.

\section{Data} \label{sec:data}
\begin{figure}
	\includegraphics[width=\columnwidth]{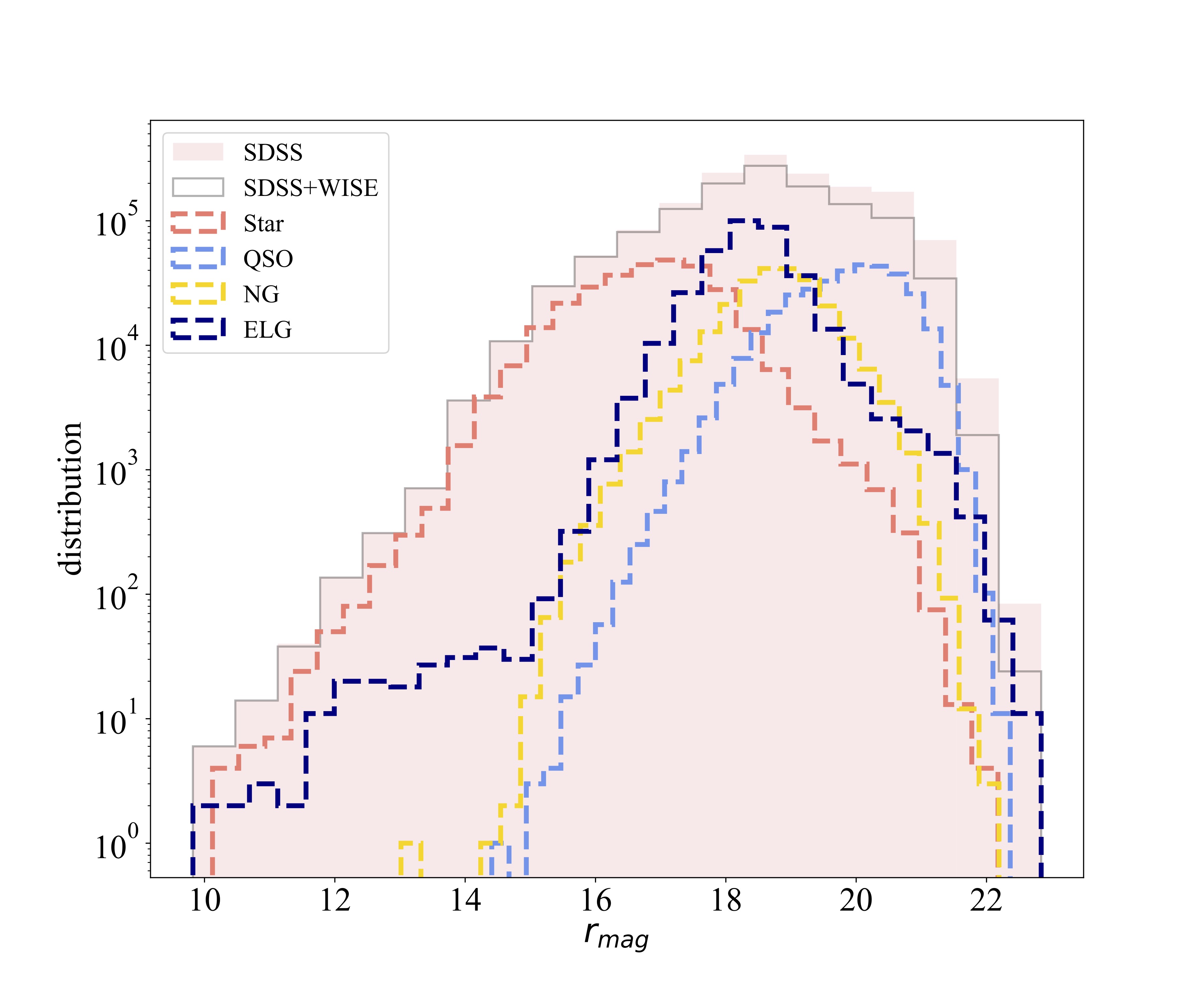}
    \caption{The $r$ magnitude distribution of data from SDSS and SDSS+ALLWISE catalogues. 
The distribution of four classes of stars, QSOs, NGs, and ELGs are plotted
in red, blue, orange, and dark blue, respectively.}
    \label{fig:distribution}
\end{figure}

We use SDSS-IV Data Release 17 (\citealt{Blanton2017, Abdurro'uf2022}) which provides a labelled dataset of spectroscopically observed sources.
SDSS mapped the sky in the five optical bands: $u$ ($\lambda = 0.355 \mu m$), $g$ ($\lambda = 0.477 \mu m$), 
$r$ ($\lambda = 0.623 \mu m$), $i$ ($\lambda = 0.762 \mu m$), and $z$ ($\lambda = 0.913 \mu m$).
The Wide-field Infrared Survey Explorer (WISE; \citealt{Wright2010}) is an all-sky survey project 
in mid-infrared band with photometry in four filters at
$W1$ ($\lambda = 3.4 \mu m$), $W2$ ($\lambda = 4.6 \mu m$), $W3$ ($\lambda = 12 \mu m$),
$W4$ ($\lambda = 22 \mu m$). 
There were hundreds of millions of celestial objects observed by the instrument, resulting in over a million images.
AllWISE has been superior to WISE in terms of deeper imaging and improved source detection.
The limiting magnitudes of $ W1_{\rm AB} $ and $ W2_{\rm AB} $ are brighter than 19.8 and 19.0 (Vega: 17.1, 15.7) 
for the AllWISE source catalogue. 
Considering the accuracy of $W1$, $W2$, $W3$, and $W4$, we only adopt $W1$ and $W2$, converting $W1$ and $W2$ in Vega magnitudes 
to AB magnitudes by $ W1_{\rm AB} = W1 + 2.699 $ and $ W2_{\rm AB} = W2 + 3.339 $ (\citealt{Schindler2017}). 
We have adopted AB magnitudes and extinction-corrected all the photometries.
Spectroscopically identified sources may be acquired by CASJOB\footnote{https://skyserver.sdss.org/casjobs/} from the SpecPhotoAll table of SDSS DR17.
Keeping the data quality, $ zWarning = 0 $, $SciencePrimary = 1 $, $ Mode = 1 $ are set when downloading data.
The records with fatal errors are rejected using flags such as $ BRIGHT $, $ SATURATED $,
$ EDGE $, and $ BLENDED $. In “where” for the SQL query,  we adopt the limitation 
as (flags \& (dbo.fPhotoFlags(`SATURATED'))) = 0 and (flags \& (dbo.fPhotoFlags(`BRIGHT'))) = 0 
and (flags \& (dbo.fPhotoFlags(`EDGE'))) = 0 and (flags \& (dbo.fPhotoFlags(`BLENDED'))) = 0.
Finally, a catalogue of 554 038 stars, 448 337 quasars, and 594 917 galaxies
which are spectroscopically identified is obtained through the query. 
We cross-match the full sample of quasars from SDSS with the Million Quasar Catalogue
(Milliquas v7.7, 2022, update \citep{Flesch2021}), so 427 829 quasars remain.
We match the SDSS-Milliquas known sources with AllWISE and the match radius is set to 3 arcsec (\citealt{Su2013}). 
The final catalogue of high quality sources with WISE photometry consists of 
305 723 stars, 345 608 quasars, and 594 077 galaxies. 
A total of two samples are drawn from the final data, Sample I is taken from SDSS+milliquas and Sample II from SDSS+milliquas+WISE.
Table~\ref{tab:table_information} shows the full information of the all 
subclasses in our samples. As for the definitions and abbreviations in Table~1, AGN is short for active galactic nucleus, 
AGN BL for broad-line AGN, SB for starburst galaxy, SB BL for broad-line SB, SF for starforming galaxy, SF BL for broad-line SF.
In the present study, the star sample is selected to the conditions: 
CLASS = STAR including all subclasses. The normal galaxy sample: CLASS = GALAXY with SUBCLASS = NULL.
AGN sample: CLASS = GALAXY with SUBCLASS = AGN and SUBCLASS = AGN BL.
Starforming sample: CLASS = GALAXY with SUBCLASS = SF and SUBCLASS = SF BL.
Starburst sample: CLASS = GALAXY with SUBCLASS = SB and SUBCLASS = SB BL.
Broad-line sample: CLASS = GALAXY with SUBCLASS = BL.
QSO sample: CLASS = QSO including all subclasses.
Fig.~\ref{fig:distribution} shows the $r$ magnitude distribution of  
our known samples from SDSS and SDSS+ALLWISE catalogues. 

In our machine learning models, we make use of the following parameters from WISE. Despite using magnitude $ \mathrm{w?mpro} $, where $ ? $ represents either $ 1 $ or $ 2 $, as measured with profile-fitting photometry (and throughout the paper, whenever we mention $ W1 $ and $ W2 $, we are actually referring to the values of $ \mathrm{w?mpro} $), we also incorporate circular aperture magnitudes in the $ W1 $ and $ W2 $ channels, denoted as $ w?mag\textunderscore n $, where $ n = 1,3 $ correspond to apertures with radii of $ 5.5\arcsec $ and $ 11\arcsec $, respectively, centered on the source \citep{Kurcz2016}.

\begin{table}
\centering
	\caption{Class and subclass distribution in the SDSS and SDSS+ALLWISE sources.}
	\label{tab:table_information}
\begin{tabular}{lllll}
\hline \hline
      & \multicolumn{1}{c}{SDSS} &         & SDSS+ALLWISE &         \\ \cline{2-2} \cline{4-4}
Class & Subclass                 & Count     & Subclass  & Count     \\ \hline
Galaxy      & AGN                      & 13 926       & AGN       & 14 305       \\
      & AGN BL                   & 2 006       & AGN BL    & 2 046       \\
      & SB                       & 51 614       & SB        & 45 817       \\
      & SB BL                    & 114       & SB BL     & 112       \\
      & SF                       & 166 678       & SF        & 170 549       \\
      & SF BL                    & 963       & SF BL     & 985       \\
      & BL                       & 10 145       & BL        & 10 387       \\
      & Null                     & 349 471       & Null      & 349 876       \\
QSO   &                          & 427 829 &           & 345 608 \\
Star  &                          & 554 038 &           & 305 723 \\ \hline
Total Count	&	&	1 576 784	&	&	1 245 408	\\ \hline
\end{tabular}
\end{table}

\section{Method} \label{sec:method}
A variety of supervised learning methods are utilized in this study, and their classification accuracy is evaluated.
In our study, we apply four machine learning models: $k$-nearest neighbors (KNN), random forest (RF; \citealt{Breiman2001}),  
XGBoost (XGB; \citealt{Chen2016}), and artificial neural network (ANN; \citealt{Haykin1998}). 
A brief introduction to each algorithm is presented as follows. 
We use Python libraries: scikit-learn \citep{Pedregosa2011}, and Keras \citep{Chollet2015}.

It is crucial to report the performance metrics of the classifiers in order to evaluate how accurate they are.
This section provides a description of the different metrics used during this research.

\subsection{Classification metrics}

Classifier performance is typically measured by Precision (P), Recall (R), and F1-score, which is a harmonic mean of Precision and Recall.
The metrics are given by:
\begin{equation}
\rm Precision = \frac{\rm TP}{\rm TP + \rm FP}
\end{equation}

\begin{equation}
\rm Recall = \frac{\rm TP}{\rm TP + \rm FN}
\end{equation}

\begin{equation}
{\rm F1{\text -}score} = 2\times{\rm Precision \times \rm Recall \over \rm Precision+\rm Recall}
\end{equation}
where TP is the number of correctly classified positives, FP the number of incorrectly 
classified positives, FN the number of incorrectly classified negatives. 

The area under the curve (AUC) of the Receiver Operating Characteristic curve (ROC; \citealt{Bradley1997}) 
is also used for the evaluation of classifier performance.
In ROC curves, true positive rate (TPR) or Recall are plotted against false positive rate (FPR) at different threshold values.

\subsection{KNN} \label{subsec:KNN}
$k$-nearest neighbors (KNN) is a classic clustering algorithm grounded on distance metrics. 
Generally, the distance can be any metric measure, and the standard Euclidean distance 
is the most common choice. Although KNN is applied to regression and classification, 
its usage is more evident as a classifier. KNN does establish a baseline of classification accuracy 
because it is the most straightforward method to use to make a classification. Specifically, 
this algorithm assumes that two data points nearby to each other falls in the same class. 
$k$ implies the number of neighboring data points to be considered and the classification 
of a source is determined by the voting results of the $k$ neighbors who are the nearest to 
the input in the multidimensional parameter space.

\subsection{Random Forest} \label{subsec:RF}
One of the machine learning (ML) methods widely used, for classification and regression tasks in astronomy 
is Random Forest (RF; \citealt{Breiman2001}). RF consists of ensembles of randomly generated 
decision trees as classifiers that uses bootstrap resampling technology. 
Bootstrapping means that several individual decision trees are trained in parallel on different 
subsets of the training dataset using randomly selected subsets of the available features.
A single decision tree tends to overfit the training data. Therefore, 
RF, which uses the average of the trees, prevents overfitting and improves prediction accuracy.
RF is fast to train and scalable, and has competitive performance to other ML algorithms.

\subsection{XGBoost} \label{subsec:XGB}
Extreme gradient boosting (XGBoost, also short for XGB) 
is an improvement of the gradient boosting algorithm \citep{Friedman2001},
a new boosting decision tree algorithm with high design efficiency, flexibility 
and strong applicability. Compared with other ensemble learning methods, 
such as the stochastic forest algorithm and the support vector machine of a single model, 
XGBoost can achieve higher performance and better generalization, has an added 
regularization term to prevent overfitting, and supports parallel computing 
(\citealt{Nguyen2019}; \citealt{Wang2021}), and has been widely used in classification and regression problems.

\subsection{ANN} \label{subsec:ANN}
Artificial neural network (ANN) mimics biological neural networks in some ways.
The network consists of an input layer, an output layer, 
and several hidden layers, each of which contains neurons that pass 
information to the neurons of the next layer. The input data 
is transmitted from the input layer through the hidden layers and reaches
the output layer where the target variable is predicted. 
In the network, the value of each neuron is a linear combination of the neurons 
in the previous layer, except the neurons in the input layer, then activated using a function that is usually non-linear.
The weights of the network are model parameters that are optimized 
by backpropagation during the training phase.
In our model, the ANN takes the photometric parameters as input.

In our study, we try two different ANN methods, one with dropout, and one without dropout, and the one with dropout overfit. 
Six ANN experiments are carried out, from a single-layer to a six-layer network. 
The experiments are conducted with two different architectures, networks composed 
of the same number of neurons in all layers, 30 neurons and 50 neurons, respectively.
The best results are obtained with a two-layer network containing 30 neurons per layer, 
a rectified linear unit (ReLu) activation is used in the hidden layer, while Softmax function 
is used in the output layer. For all our hyper-parameter sweeps, we train the model for 
1000 epochs for the Adams optimizer and sparse categorical crossentropy as the loss function. 
We use the early stopping technique to prevent the model from overfitting the training data. 

\subsection{Voting} \label{subsec:voting}
Voting is an ensemble ML algorithm that combines the output of several
ML models and often shows better results than a single model, and it can be used for regression or classification.
For classification, the final result is determined by incorporating the output
of multiple models using soft or hard voting. 
A hard-voting ensemble aggregates the votes for class labels 
from other classifiers and predicting the class with the most votes. 
A soft-voting ensemble is calculated on the predicted probability of the output class.
This study uses a voting classifier 
where the ensemble comprises three classifiers.
As different base classifiers contribute differently to the final classification result,
the classifiers with better performance are weighted more heavily. Here, the weights 
for RF and XGB are 2 and for KNN is 1 as the former two classifiers perform better
than the latter one.

\subsection{Model Setup} \label{subsec:model_setup}
We adopt two main parts in the methodology used in this work. 
The experiments are conducted to test the ability to classify objects into three, four, and seven different categories.
The three-class experiment includes star, QSO, and galaxy; the four-class 
experiment includes star, QSO, NG, and ELG;
and the seven-class experiment includes star, QSO, NG, AGN, SF, SB,
and BL.
As we complete three experiments, we explore the possibility of classifying galaxies 
into normal and emission-line ones, and then a more detailed classification of emission-line galaxies
into AGN, SF, SB, and BL galaxies.
In the following, the process by which the two parts of this study are carried out will be described in detail.

\subsubsection{The first part} \label{subsubsec:first step}
\begin{figure*}
	\includegraphics[width=18cm]{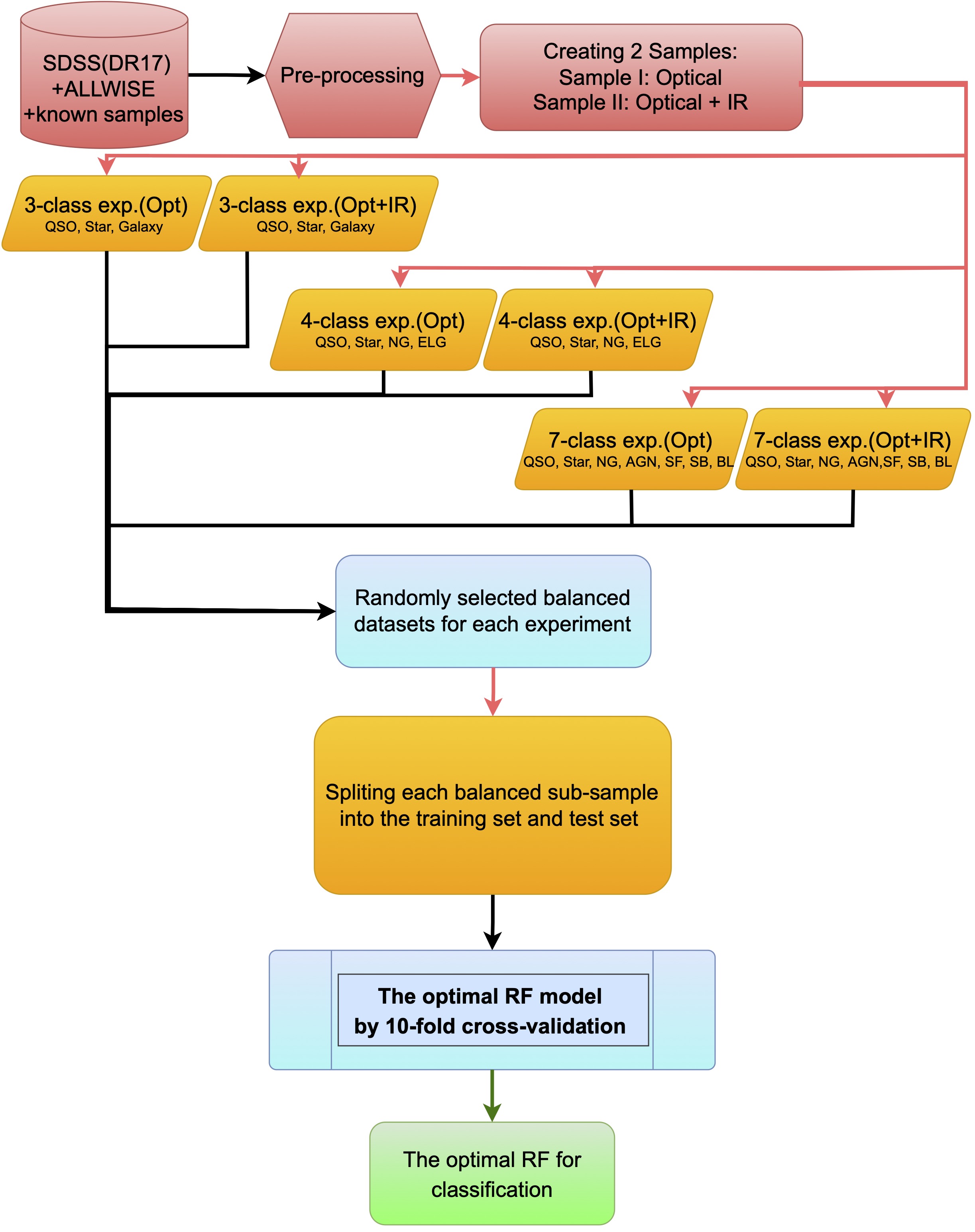}
    \caption{This flowchart outlines the first part of our methodology, 
using color-coded shapes and arrows to represent the dataset, 
experiments, functions, and results. Red and orange shapes 
indicate the raw and processed data, respectively, while blue shapes 
represent functions performed. Green shapes signify the output. 
Arrows are color-coded to indicate input (black), data splitting (red), and output (green).}
    \label{fig:flowchart}
\end{figure*}

As the first part of our approach shown in the flowchart of Fig.~\ref{fig:flowchart}, we perform the experiments using the RF method 
in order to work out a reliable classification of astronomical objects.
The experiments are three-, four-, and seven-class experiments in which the classifier is trained on the balanced datasets. 
All the experiments are conducted on both Samples I and II, including optical and optical+IR data, respectively.
The number of data used in three experiments is as follows.

\emph{The three-class experiment} classifies the sources into stars, QSOs, and galaxies.
The randomly selected sources of each class for training are 420 000 and 305 000 from 
Samples I and II, respectively.

\emph{The four-class experiment} classifies the sources into stars, QSOs, 
NGs, and ELGs.
The number of each class selected randomly for the training dataset is
245 000 and 244 000 from Samples I and II, respectively.

\emph{The seven-class experiment} classifies the sources into stars, QSOs, 
NGs, AGNs, SFs, SBs,
and BLs. The number of randomly selected sources for each class 
from either sample is 10 000.

\subsubsection{The second part} \label{subsubsec:second step}
In the second part, which actually comes from the best results 
of the first part, we classify the objects into three and four classes.
We train several different ML methods: KNN, RF, XGB, voting, 
and ANN on both Samples I and II and used all the data we have.
It should be noted that in part one, we test our model on a balanced 
dataset in order to ensure that it performs well in different experiments. 
Therefore, in part two, we use all the existing data that we already have.
Following is the number of data in each experiment.

\emph{Three-class classification using imbalanced datasets:} the two datasets used 
in the first experiment for the three-class classification are as follows:
for Sample I (optical dataset), star: 554 034, QSO: 427 830, galaxy: 594 917; for Sample II (optical+IR dataset), star: 305 719, QSO: 345 608, galaxy: 594 077. 

\emph{Four-class classification using imbalanced datasets:} the two datasets used 
in the second experiment for the four-class classification are as follows:
for Sample I (optical dataset),  star: 554 034, QSO: 427 830, ELG: 245 446, and NG: 349 471;
for Sample II (optical+IR dataset), star: 305 719, QSO: 345 608, ELG: 244 201, and NG: 349 876.

For all ML methods, their purposes are to construct models and assess their proficiency 
on previously unseen data. To obtain an average of all evaluation metrics, 
we employ a 10-fold cross-validation approach. This iterative method randomly 
splits the sample into ten portions, where nine sections are used for training 
and one part is reserved for test and performance calculation in each iteration. 
We repeat this process ten times, resulting in a distinct test set for each iteration. 
This procedure is essential in determining the model's performance. 
Finally, we average the ten performance metric estimates.

\section{Result} \label{sec:result}
We present the results of all experiments that are carried out 
in the two parts of this study in this section.

\subsection{Feature selection}
\begin{figure*}
    \centering
    \begin{tabular}{ccc}
        \begin{subfigure}{0.3\textwidth}
            \centering
            \includegraphics[width=\linewidth]{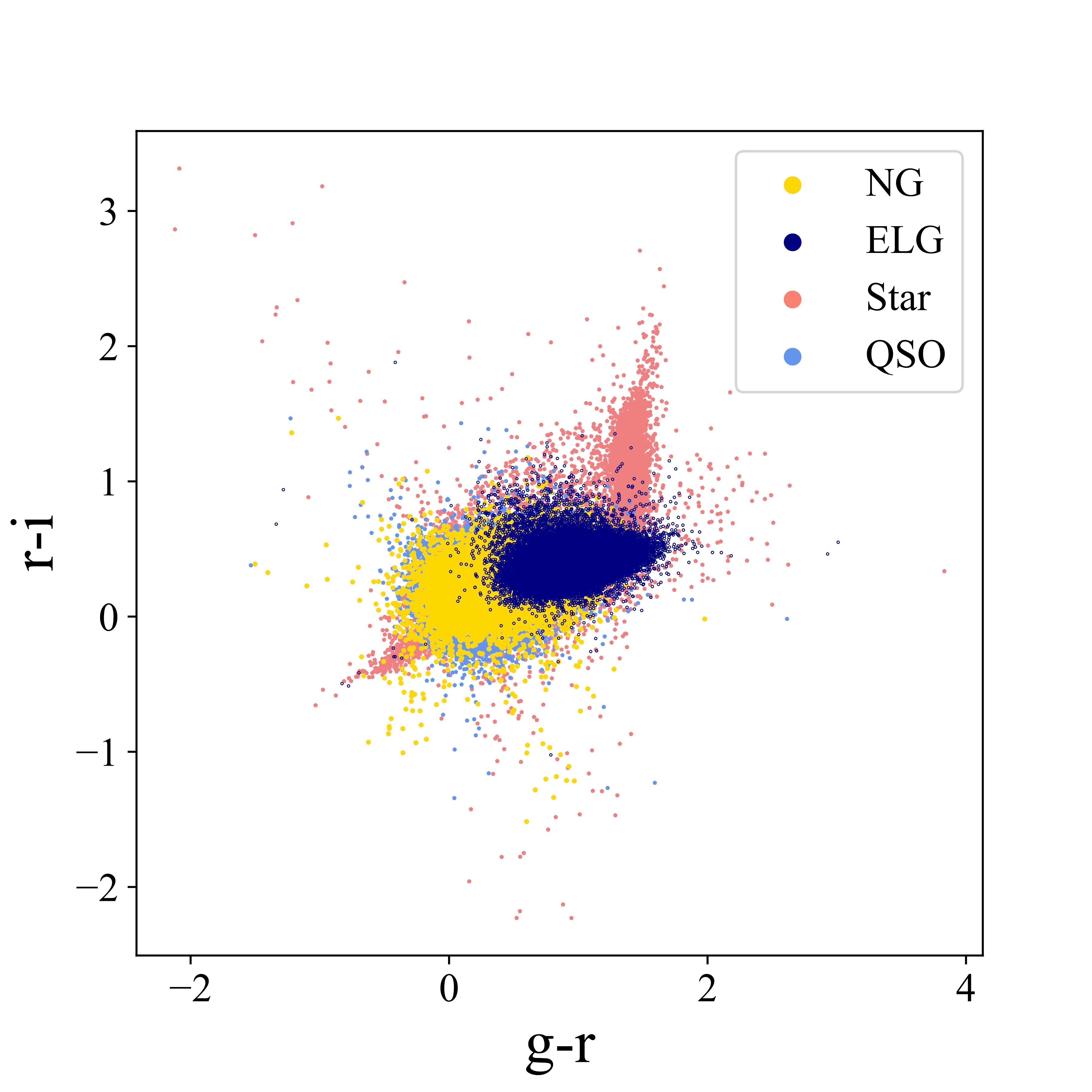}
            \label{fig:1}
        \end{subfigure} &
        \begin{subfigure}{0.3\textwidth}
            \centering
            \includegraphics[width=\linewidth]{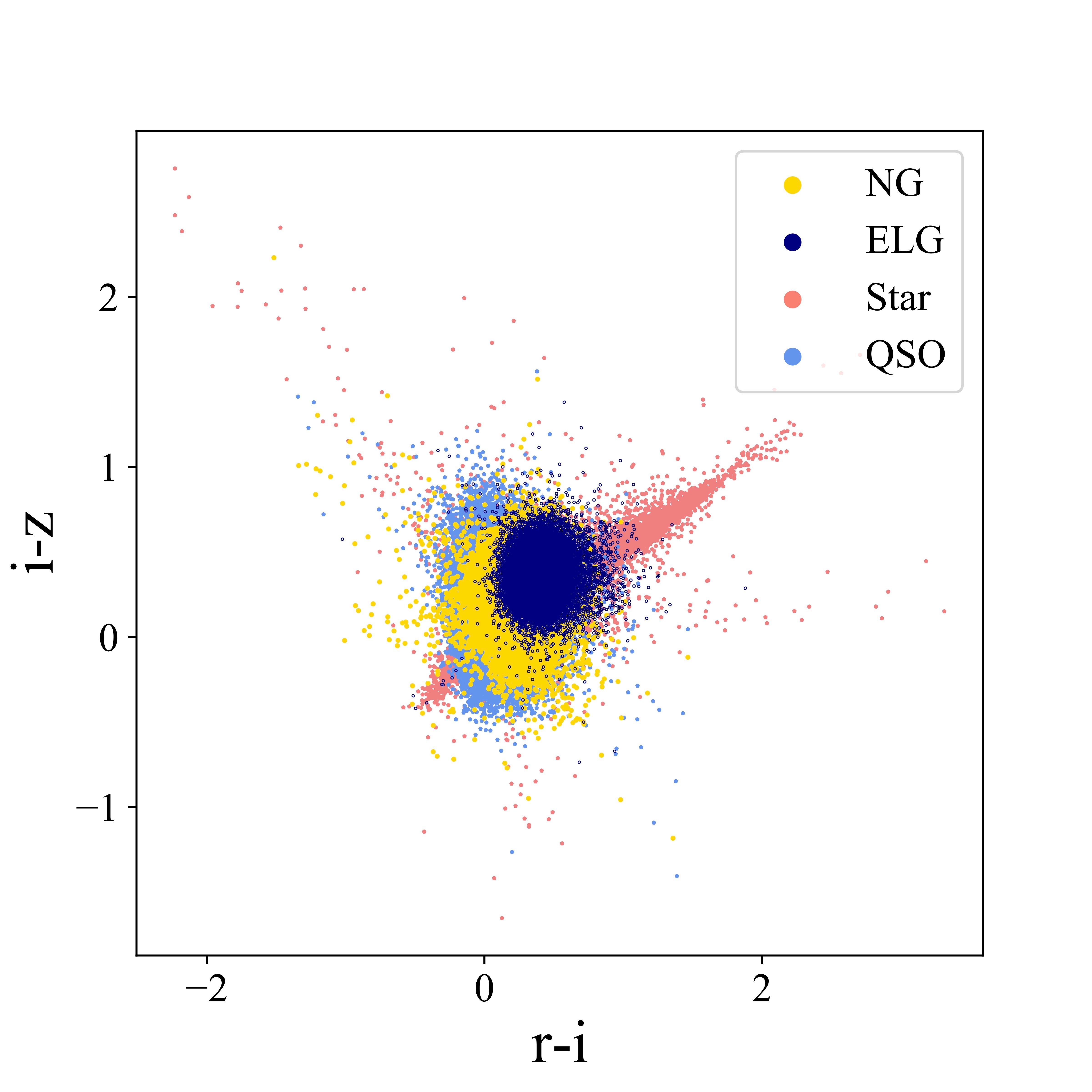}
            \label{fig:2}
        \end{subfigure} &
        \begin{subfigure}{0.3\textwidth}
            \centering
            \includegraphics[width=\linewidth]{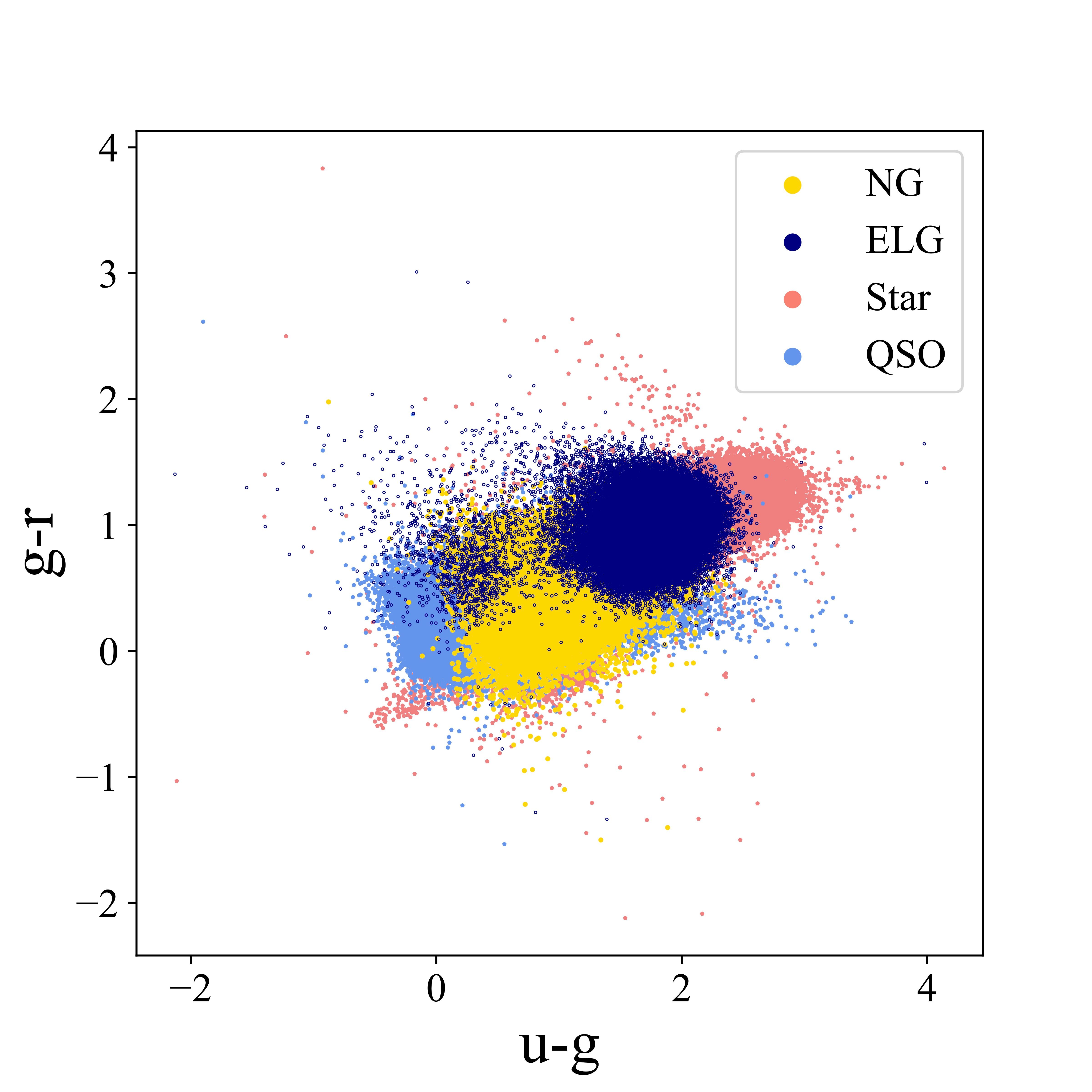}
            \label{fig:3}
        \end{subfigure} \\
        
        \begin{subfigure}{0.3\textwidth}
            \centering
            \includegraphics[width=\linewidth]{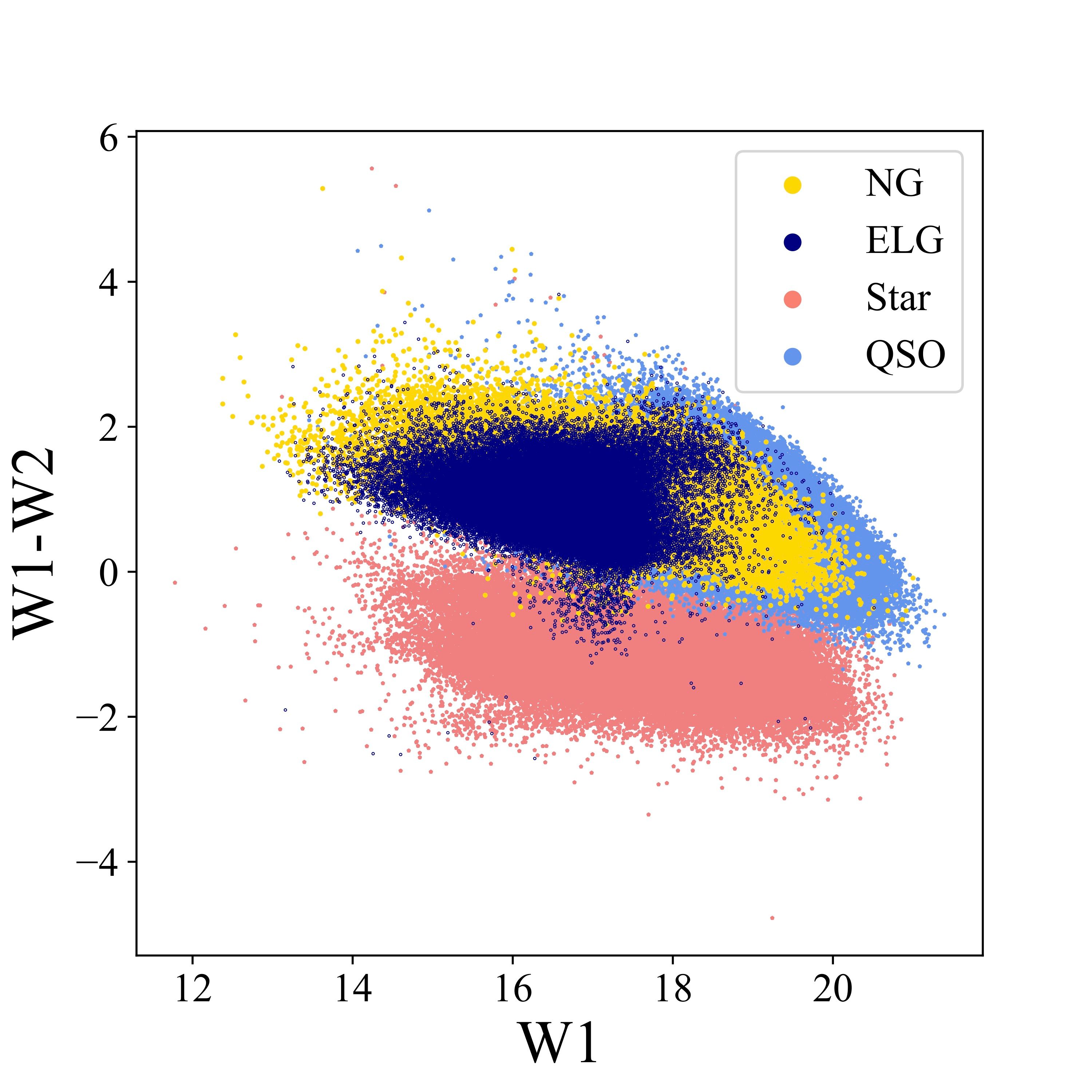}
            \label{fig:4}
        \end{subfigure} &
        \begin{subfigure}{0.3\textwidth}
            \centering
            \includegraphics[width=\linewidth]{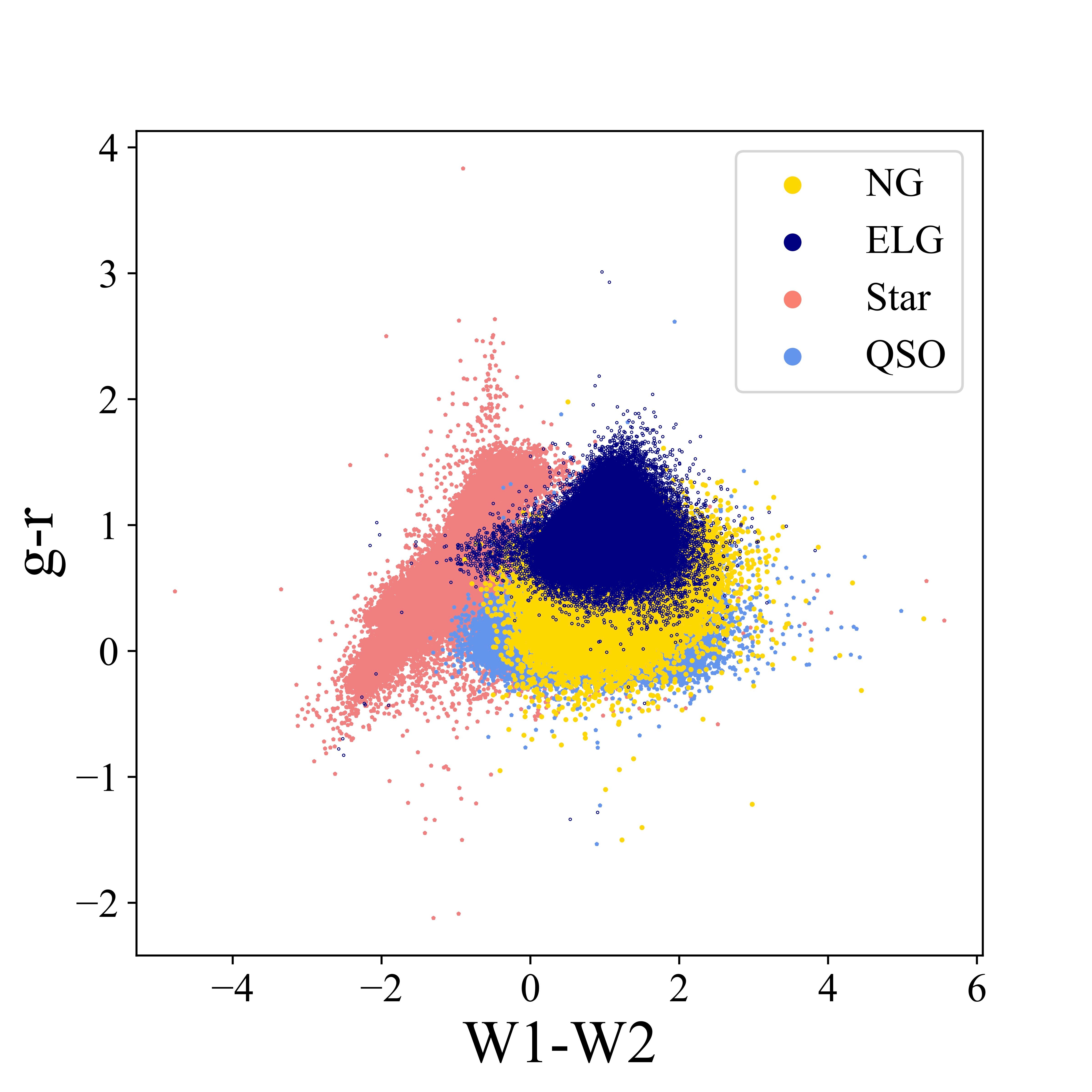}
            \label{fig:5}
        \end{subfigure} &
        \begin{subfigure}{0.3\textwidth}
            \centering
            \includegraphics[width=\linewidth]{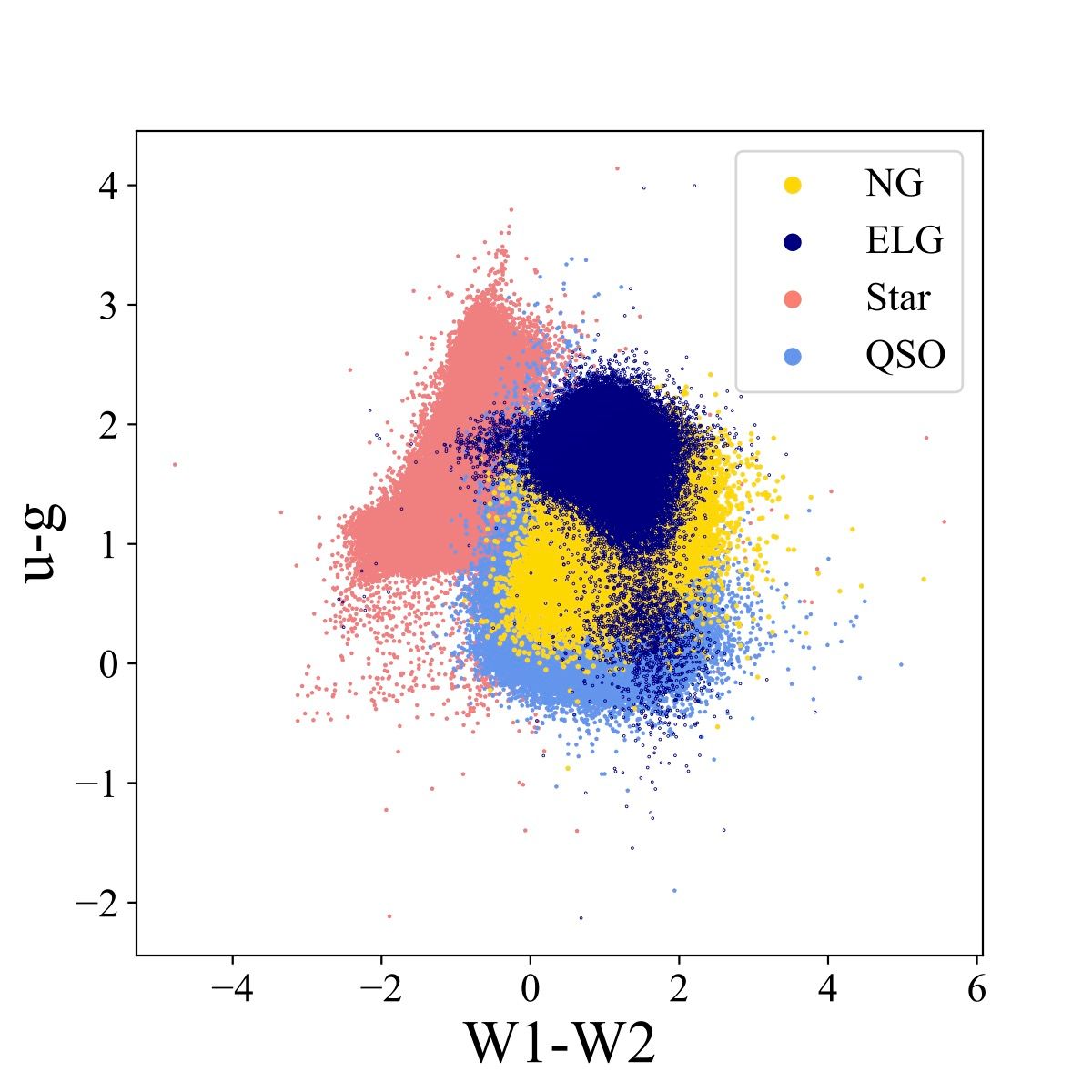}
            \label{fig:6}
        \end{subfigure} \\       
    \end{tabular}    
    \caption{Color-color plots showcasing the classification of four classes of astronomical objects,  
stars, quasars (QSOs), normal galaxies (NGs), and emission line galaxies (ELGs) using optical and infrared data. 
The inclusion of WISE (or IR) information enhances accuracy in distinguishing between stars, quasars, and galaxies. 
}
    \label{fig:color-color}
\end{figure*}

To assess the significance of infrared and optical data for different object categories, we have included color-color plots in Fig.~\ref{fig:color-color}. These plots effectively demonstrate the influence of infrared data in distinguishing stars from other astronomical objects. Solely relying on optical information tends to result in overlap between stars and other objects, as evident in the first row of Fig. \ref{fig:color-color}. However, the inclusion of extended information such as infrared data facilitates improved differentiation between active and normal galaxies and pointed sources such as stars and quasars, as observed in the second row of the figure. Notably, the role of infrared information is particularly critical for accurately classifying stars and quasars.

Using a larger parameter space, as opposed to a simple color-color-based classification, offers significant advantages. Although simple color selections in the optical or infrared are effective for identifying certain object classes, a larger parameter space enables a more detailed and nuanced classification. It allows the algorithm to capture a broader range of object characteristics and consider additional features beyond color information. Consequently, this leads to a more refined classification scheme and improved discrimination between different object categories, especially when the boundaries between classes are less distinct.

A classifier's performance can be determined by what features influence it the most.
In the classification process, we choose features from the literature where the most popular ML models perform well \citep{Li2021}.
As we have two samples with optical and infrared features, we select two feature sets: $ r, u-g, g-r, r-i, i-z $ for Sample I, and $ r, u-g, g-r, r-i, i-z, z-W1, W1-W2 $ for Sample II.
We have also employed two additional sets of input features, namely including difference of two circular aperture magnitudes, $ r, u-g, g-r, r-i, i-z, z-W1, W1-W2, w1mag1-w1mag3 $ and $ r, u-g, g-r, r-i, i-z, z-W1, W1-W2, w1mag1-w1mag3, w2mag1-w2mag3 $. 
These are used solely for the purpose of evaluating the influence (or lack thereof) of apertures on the classification of astronomical objects in the context of the four-class classification experiment.

\begin{figure*}
\centering
        \begin{subfigure}[b]{0.5\textwidth}
                \centering
                \includegraphics[width=\linewidth]{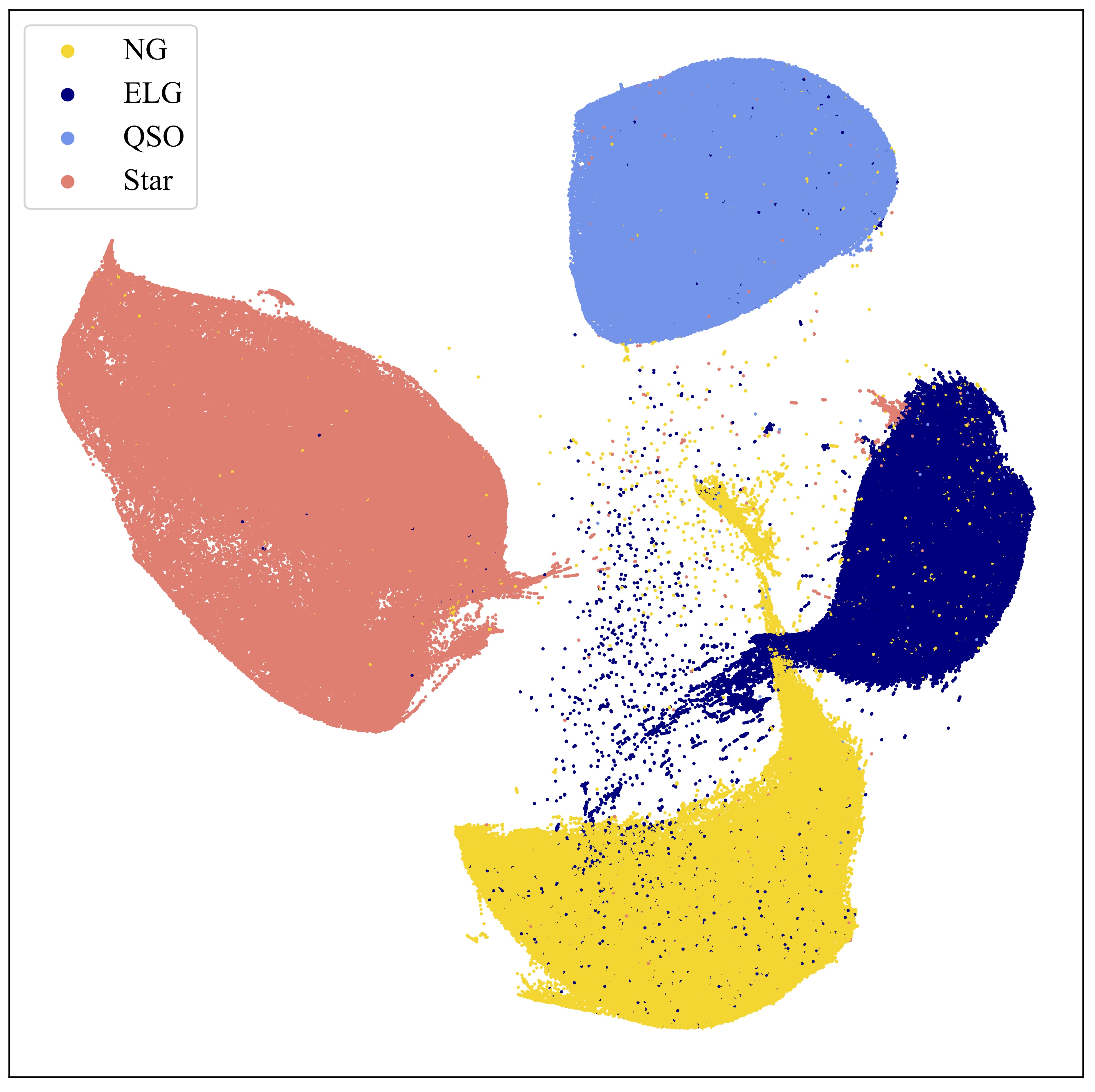}
                %\caption{}
                \label{fig:QSO}
        \end{subfigure}\hfill
        \begin{subfigure}[b]{0.5\textwidth}
                \centering
                \includegraphics[width=\linewidth]{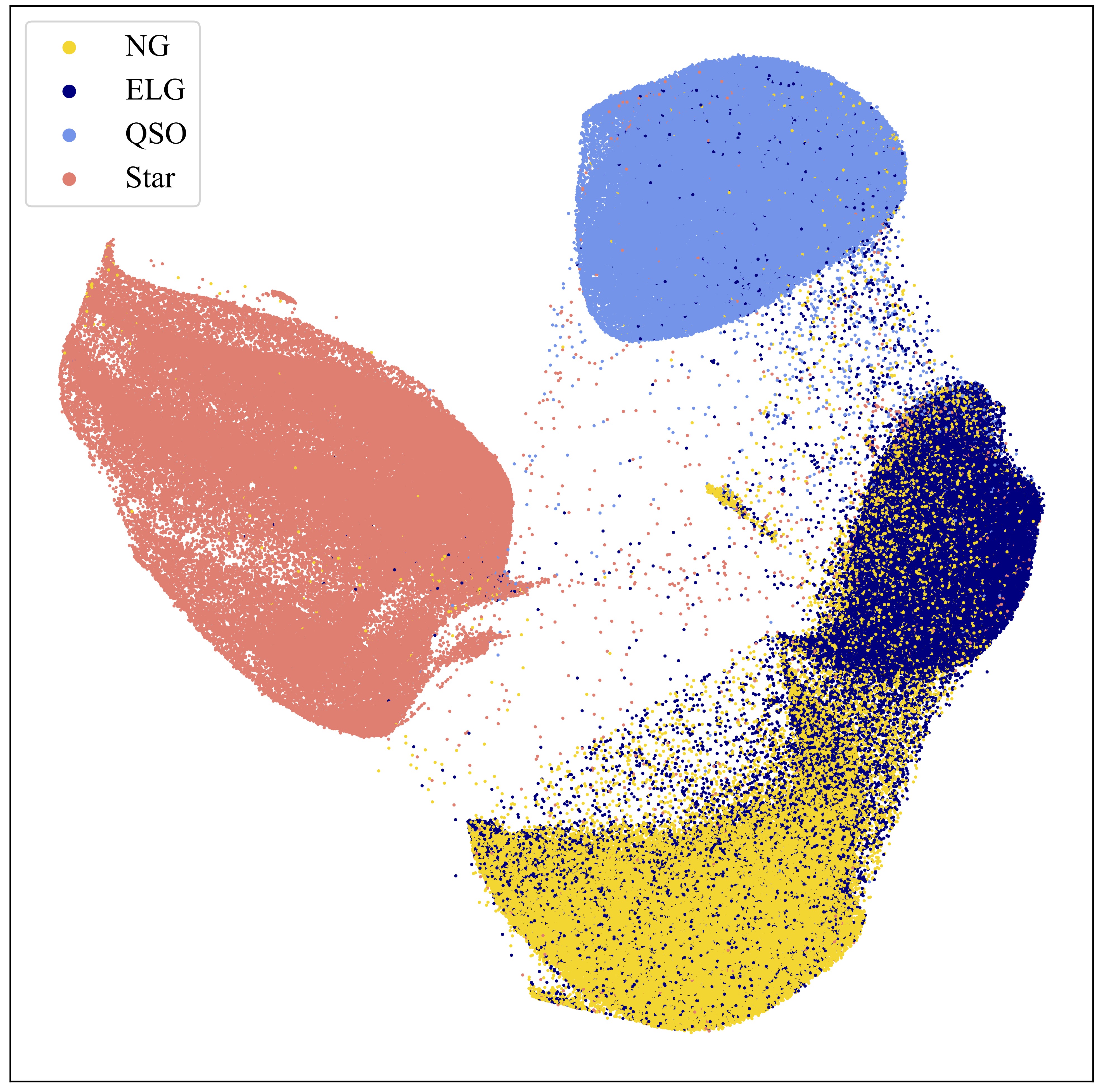}
                %\caption{s}
                \label{fig:NG}
        \end{subfigure}
        \caption{Visualizing the data points using supervised UMAP for four distinct classes, including normal galaxies (NG), emission line galaxies (ELG), quasars (QSO), and stars. The left panel represents the training set, while the right panel displays the test set. The input features utilized for classification are u-g, g-r, r-i, i-z, z-W1, and W1-W2.}\label{fig:umap}
\end{figure*}

In our study, we employ supervised Uniform Manifold Approximation and Projection (UMAP) visualization, as depicted in Fig.~\ref{fig:umap}, to enhance our comprehension of the distribution and distinguishability of celestial objects using the features we have extracted. The input features utilized for classification are $ u-g, g-r, r-i, i-z, z-W1 $, and $ W1-W2 $. This visualization serves as a powerful tool for understanding the underlying structure of our data and, notably, for assessing the efficacy of our feature selection process. By observing the UMAP plot, which presents four-class classification of NGs, ELGs, QSOs, and Stars, we can identify distinct clusters or groupings of data points that correspond to different object types. The left panel of the UMAP plot displays the training data, while the right panel shows the test data. This emphasizes the success of our feature selection and machine learning approach in achieving clear separation between various object classes, a crucial aspect of our methodology. Additionally, the UMAP visualization aids in recognizing any disparities or overlaps between training and test datasets, providing useful information for fine-tuning our model and enhancing its generalizability. Such insights are instrumental in optimizing our feature selection strategy for robust classification performance.

%%%%%%%
\begin{figure}
   \begin{minipage}{0.48\textwidth}
     \centering
     \includegraphics[width=1\linewidth]{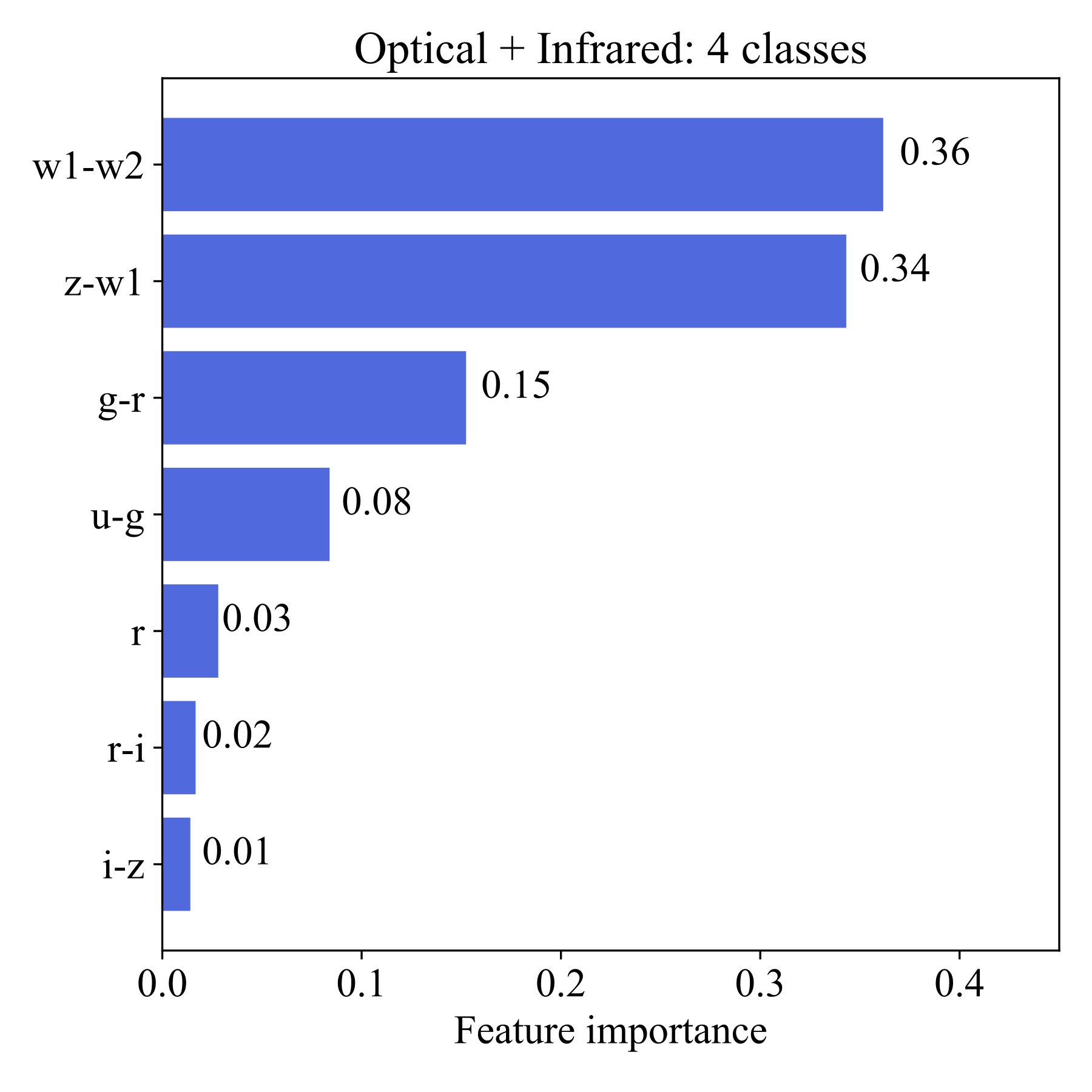}
     \caption{Feature importance of XGB model with optical and infrared information 
                  for Sample II in the four-class experiment}
\label{fig:feature_importance}
   \end{minipage}
\end{figure}

\begin{figure}
 \begin{minipage}{0.5\textwidth}
     \centering
     \includegraphics[width=1\linewidth]{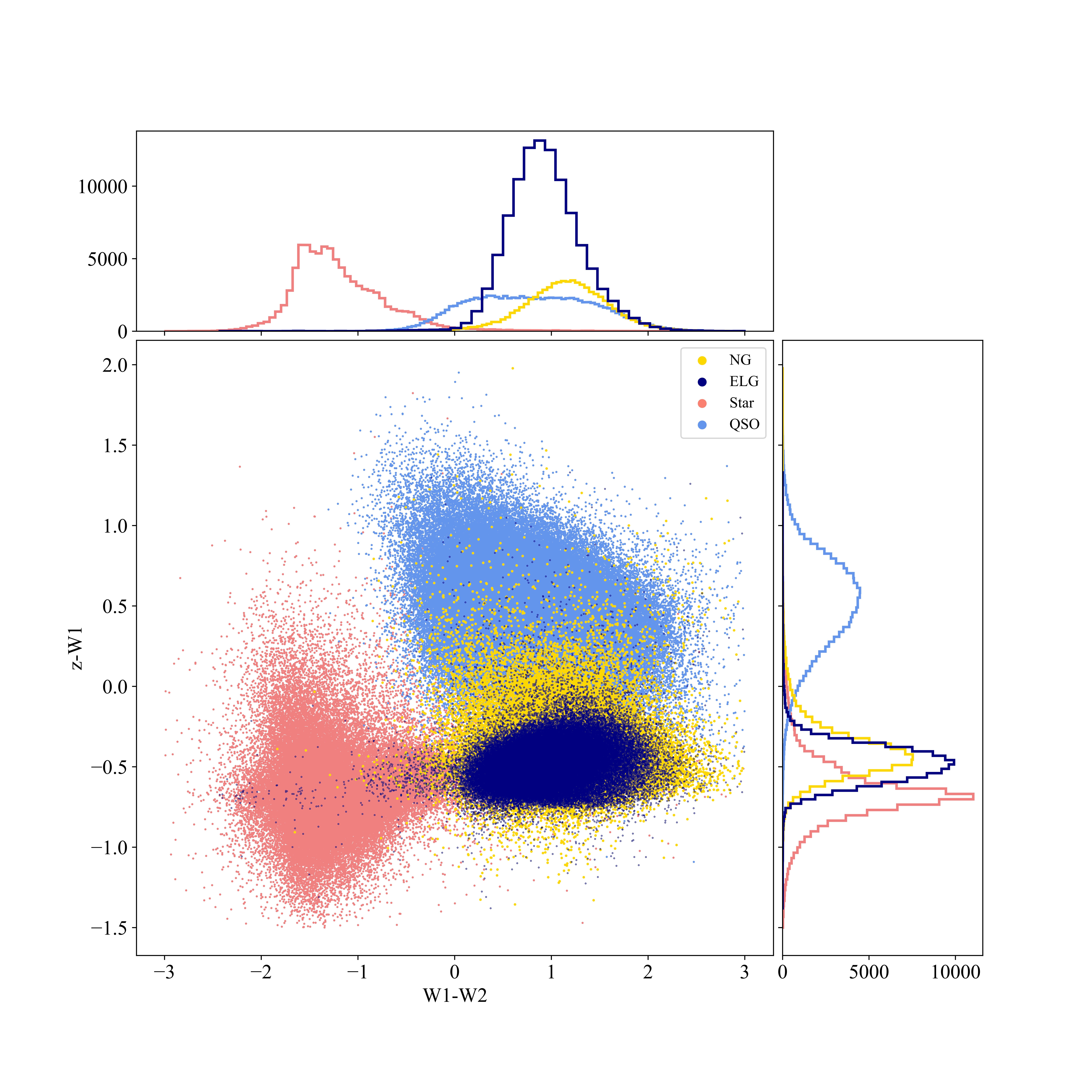}
     \caption{Color-Color distribution of NGs, ELGs, stars, and QSOs based on the most important features in XGB classifier. The data are from Sample II with optical and infrared information.} 
\label{fig:color_dist}
   \end{minipage}\hfill
\end{figure}

Building on the effectiveness of our feature selection and machine learning approach, particularly the superiority of XGB over other algorithms, we proceed to assess feature importance using Fig.~\ref{fig:feature_importance}.
The most important features in the figure are the infrared features, $ W1-W2 $ with an importance 
of 36 per~cent, and the combination of optical and infrared features, $ z-W1 $ with an importance of 34 per~cent.
Based on these two features that have the most influence on classification, we plot Fig.~\ref{fig:color_dist}. 
In this figure, the pair of features as well as color distribution shows 
the differences among the four classes of NG, ELG, star, and QSO. 
As shown in Fig.~\ref{fig:color_dist}, it is obvious that stars, 
quasars and galaxies (including normal and emission-line galaxies) 
are easy to separate, especially stars, while normal galaxies and emission-line galaxies overlap seriously.

\subsection{Classification results of the first part}
The results of the evaluation metrics, Precision, Recall, F1, and AUC  
for three-, four-, and seven-class experiments are presented in Table~\ref{tab:RF}.
At first glance, we can see adding the infrared features improves the evaluation metrics in all experiments.
For Sample II,  in the three-class experiment, the RF classifier performs well, for example, F1 of 
galaxy, QSO, and star are 0.986, 0.992, and 0.988, respectively.

\begin{table}
\centering
	\caption{Performance comparison of RF classifier in the experiments for Samples I and II with optical and infrared information:
	Comparison of boldface results with literature in Section \ref{sec:discussion}.}
	\label{tab:RF}
\begin{tabular}{cccccc}
\hline
\hline
%\cline{1-6}  \\
                                  &  Class                & P & R & F1    & AUC   \\
%\cline{1-6}  \\
\hline
%                           &       & \multicolumn{1}{l}{Three-class} &        &       &                 \\ \\
\multirow{3}{*}{Optical}          & Galaxy               & 0.945     & 0.934  & 0.939 & 0.985 \\
                                  & Star                 & 0.921     & 0.939  & 0.93  & 0.987 \\
                                  & QSO                  & 0.944     & 0.937  & 0.94  & 0.988 \\
\hline
\multirow{3}{*}{Optical+IR} & Galaxy               & 0.987     & 0.985  & \textbf{0.986} & 0.994 \\
                                  & Star                 & 0.984     & 0.993  & \textbf{0.988} & 0.995 \\
                                  & QSO                  & 0.996     & 0.989  & \textbf{0.992} & 0.994 \\
\hline
%  \\ \cline{1-6} \\  
%                                    &       & \multicolumn{1}{l}{Four-class} &        &       &                  \\ \\
\hline
\multirow{4}{*}{Optical}          & ELG & 0.82      & 0.834  & 0.827 & 0.959 \\
                                  & NG        & 0.775     & 0.748  & 0.761 & 0.933 \\
                                  & Star                 & 0.929     & 0.926  & 0.928 & 0.987 \\
                                  & QSO                  & 0.9       & 0.92   & 0.91  & 0.986 \\ 
\hline
\multirow{4}{*}{Optical+IR} & ELG & 0.853     & 0.856  & 0.854 & 0.971 \\
                                  & NG        & 0.847     & 0.84   & 0.843 & 0.968 \\
                                  & Star                 & 0.995     & 0.987  & 0.991 & 0.994 \\
                                  & QSO                  & 0.977     & 0.99   & 0.983 & 0.995 \\
%  \\ \cline{1-6} \\  
%         &       & \multicolumn{1}{l}{Seven-class} &        &       &                  \\ \\
\hline\hline
\multirow{7}{*}{Optical}          & AGN                  & 0.43      & 0.398  & 0.414 & 0.84  \\
                                  & SF          & 0.545     & 0.53   & 0.537 & 0.883 \\
                                  & SB            & 0.699     & 0.722  & 0.71  & 0.948 \\
                                  & BL            & 0.593     & 0.64   & 0.616 & 0.918 \\
                                  & Star                 & 0.899     & 0.899  & 0.899 & 0.987 \\
                                  & QSO                  & 0.846     & 0.871  & 0.858 & 0.982 \\
                                  & NG               & 0.506     & 0.485  & 0.495 & 0.869 \\  
\hline
\multirow{7}{*}{Optical+IR} & AGN                  & 0.551     & 0.499  & 0.524 & 0.88  \\
                                  & SF          & 0.635     & 0.635  & 0.635 & 0.923 \\
                                  & SB            & 0.787     & 0.825  & 0.805 & 0.976 \\
                                  & BL            & 0.633     & 0.659  & 0.646 & 0.93  \\
                                  & Star                 & 0.993     & 0.981  & 0.987 & 0.993 \\
                                  & QSO                  & 0.949     & 0.974  & 0.961 & 0.994 \\
                                  & NG               & 0.596     & 0.588  & 0.592 & 0.908 \\ 
\hline\hline
\end{tabular}
\end{table}

In the four-class experiment, the RF classifier still works well. 
The F1 for star, QSO, NG, and ELG are 0.991, 0.983, 0.843, and 0.854, respectively.
The AUC scores for star, QSO, NG, and ELG are 0.994, 0.995, 0.968, and 0.971, respectively.

\begin{figure}
   \begin{minipage}{0.48\textwidth}
     \centering
     \includegraphics[width=.9\linewidth]{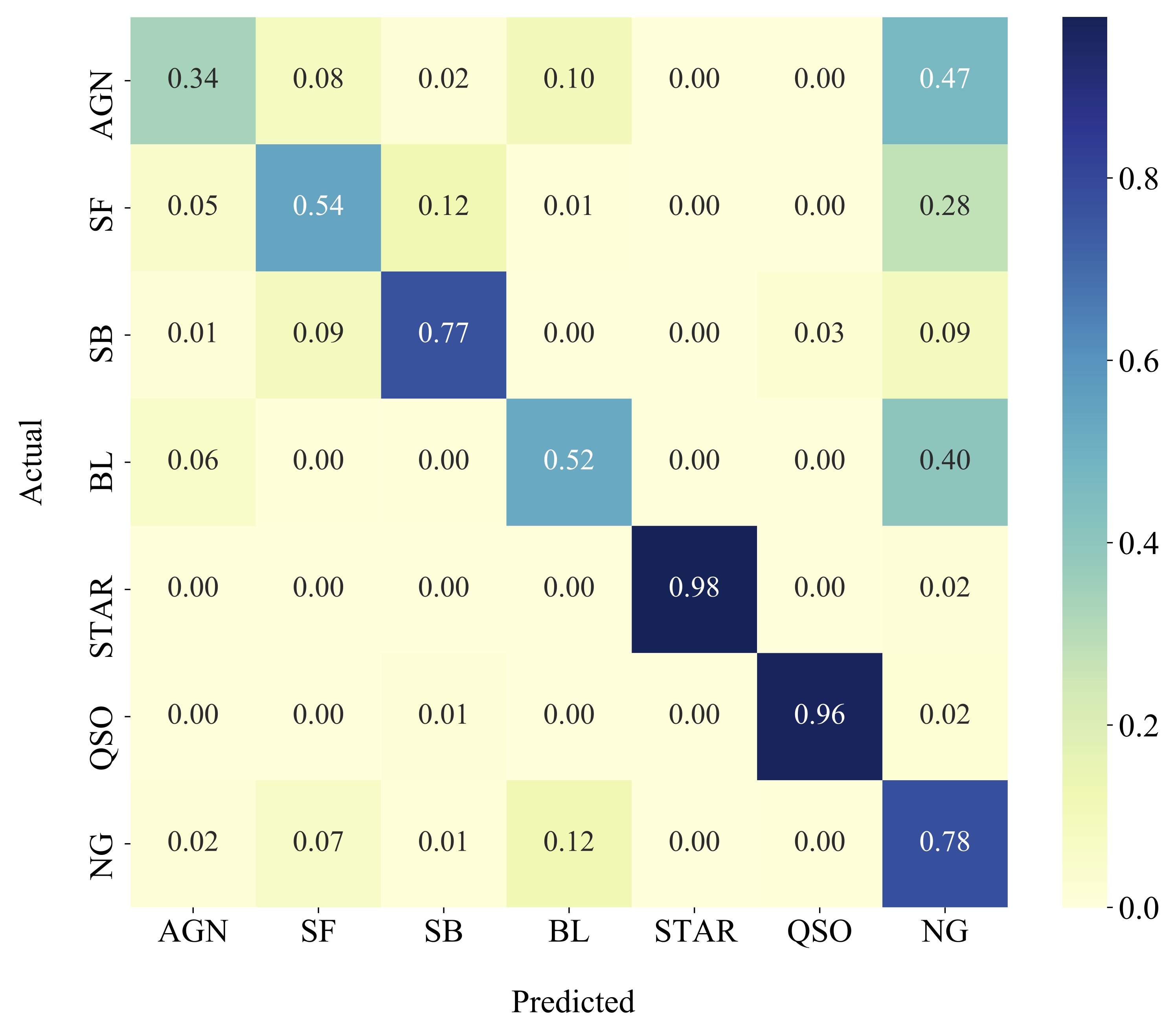}
     \caption{Confusion matrix of RF classifier in the seven-class experiment for Sample II.} \label{fig:confusion}
   \end{minipage}\hfill
   \begin{minipage}{0.48\textwidth}
     \centering
     \includegraphics[width=.8\linewidth]{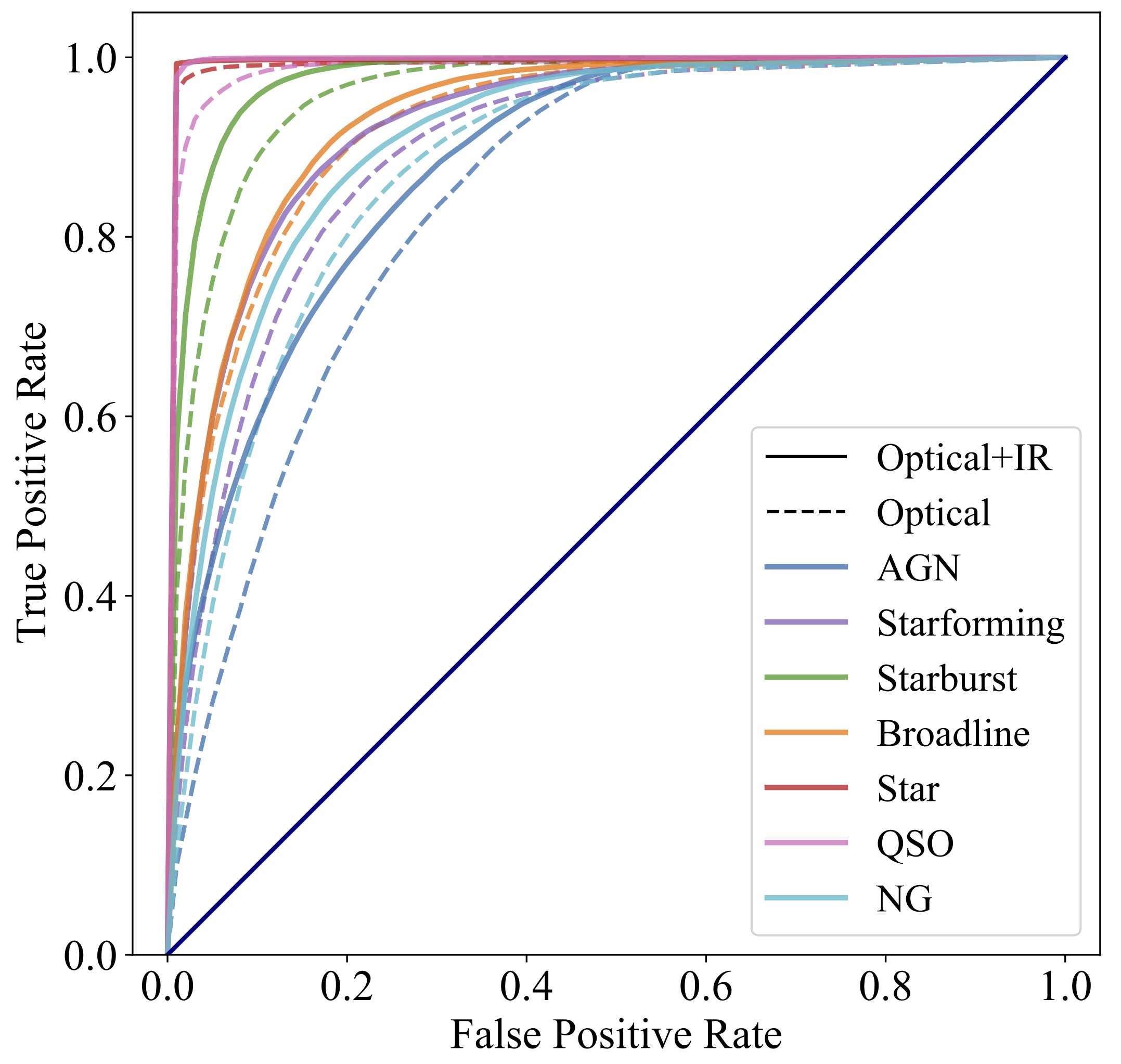}
     \caption{The ROC curve of RF classifier in the seven-class experiment. 
The solid and dashed lines, respectively, show the ROC curves of Samples I and II with optical and infrared information.
}\label{fig:roc curve}
   \end{minipage}
\end{figure}

As a result of the seven-class experiment, F1 is 0.987 for stars, 0.961 for quasars, 
while F1 for the other classes, including AGN, SF, SB, BL, and NG are 0.524, 0.635, 0.805, 0.646, 0.592, respectively.
The confusion matrix of the seven-class experiment of Sample II
is shown in Fig.~\ref{fig:confusion}. The vertical square boxes are actual labels
and the horizontal boxes show the RF classifiers' predictions. In the matrix, 
more populated areas are shown with darker colors. As can be seen from the confusion matrix,
47 per~cent of AGNs, 28 per~cent of SFs, 10 per~cent of SBs, and 41 per~cent of BLs are misclassified as NG. 
Fig.~\ref{fig:roc curve}, the AUC of the RF classifier for the seven-class experiment,
also confirms the results shown in the confusion matrix. 
A perfect classifier has TPR = 1 and FPR = 0, resulting in an AUC score of 1. 
Fig.~\ref{fig:roc curve} shows that the stars have the highest AUC score, 
making them the best-predicted class, while the AGN class has the lowest AUC score.
As a result of the high percentage of misclassification of three classes of AGN, SF, and BL 
with NGs, we decide to do the four-class experiment with NG, ELG, star, 
and QSO using several different ML methods in the second part of our work.
This means that we have merged four classes of AGN, SF, SB, and BL 
into ELGs as an independent class. As a continuation of this section, 
we present the results of experiments conducted on three- 
and four-class in the second part of our approach and compare the performance of the classifiers.

%%%%%%

\subsection{Effect of aperture magnitudes on classification performance}
We conduct experiments that involve the incorporation of differential aperture magnitudes, specifically $ w1mag1-w1mag3 $ or $ w1mag1-w1mag3 $ and $ w2mag1-w2mag3 $, the results of which are presented in Table~\ref{tab:aperture}.
These experiments are exclusively carried out for the four-class classification to evaluate the influence of different aperture magnitudes on our classification process.  We compare the results obtained when including the differential aperture magnitudes to those when excluding them.
The findings from our experiments reveal that the F1-score for stars and QSOs remains unchanged, while there is a 4 per~cent improvement in the F1-score for NGs, increasing from 0.84 to 0.88. However, the F1-score for ELGs shows a 4 per~cent decrease, dropping from 0.85 to 0.81.
Astronomers typically strive for high-performance classification of emission line galaxies. Given the observed decline in performance for this category, we have opted not to include these parameters in our subsequent experiments.

\subsection{Classification results of the second part: comparison of the different ML models}
In this subsection, the results of the experiments performed in 
Sec.~\ref{subsubsec:second step} for both Sample I and Sample II 
are represented. The performance metrics (Precision, Recall and F1) of 
the different ML models, KNN, RF, XGB, voting, and ANN in three- 
and four-class experiments are shown in Tables~\ref{tab:three-class} 
and \ref{tab:four-class}, respectively. In the three-class experiment in 
Table~\ref{tab:three-class}, the evaluation metrics of all classifiers for galaxy class in 
Sample II (optical+IR) perform similarly; RF, XGB, and 
voting have the F1 values of 0.99; KNN and ANN have the values of 0.989.
XGB has the best performance for both star and QSO with the F1 values of 0.986 and 0.992, respectively.

\begin{table*}
\centering
	\caption{Performance comparison of ML models in the three-class experiment for Samples I and II with optical and infrared information.
	Comparison of boldface results with literature in Section \ref{sec:discussion}.}
	\label{tab:three-class}
\begin{tabular}{c|c|ccc|ccc|ccc}
\hline
\multicolumn{1}{c}{} & \multicolumn{3}{c}{Galaxy} & \multicolumn{3}{c}{Star} & \multicolumn{3}{c}{QSO}                    \\
\multicolumn{1}{c}{Features} & Method & P       & R       & F1      & P       & R      & F1     & P     & R     & \multicolumn{1}{c}{F1}    \\ 
\hline
\multicolumn{1}{c}{\multirow{5}{*}{Optical}}          & KNN    & 0.945 & 0.930    & 0.937   & 0.919   & 0.938  & 0.928  & 0.941 & 0.935 & \multicolumn{1}{c}{0.938} \\
\multicolumn{1}{c}{}                                  & RF        & 0.949   & 0.940    & 0.945   & 0.912   & 0.924  & 0.918  & 0.944 & 0.944 & \multicolumn{1}{c}{0.944} \\
\multicolumn{1}{c}{}                                  & XGB    & 0.950    & 0.940   & 0.945   & 0.913   & 0.926  & 0.920   & 0.944 & 0.944 & \multicolumn{1}{c}{0.944} \\
\multicolumn{1}{c}{}                                  & Voting  & 0.947   & 0.933   & 0.940    & 0.921   & 0.940   & 0.931  & 0.944 & 0.938 & \multicolumn{1}{c}{0.941} \\
\multicolumn{1}{c}{}                                  & ANN    & 0.947   & 0.932   & 0.939   & 0.927   & 0.932  & 0.929   & 0.931 & 0.940  & \multicolumn{1}{c}{0.936} \\ \hline
\multicolumn{1}{c}{\multirow{5}{*}{Optical+IR}} & KNN    & 0.988  & 0.989   & 0.989   & 0.981   & 0.989  & 0.985  & 0.996 & 0.984 & \multicolumn{1}{c}{0.990}  \\
\multicolumn{1}{c}{}                                  & RF       & 0.990    & 0.989   & \textbf{0.990}    & 0.981   & 0.990   & \textbf{0.985}  & 0.995 & 0.987 & \multicolumn{1}{c}{\textbf{0.991}} \\
\multicolumn{1}{c}{}                                  & XGB    & 0.991   & 0.989   & \textbf{0.990}    & 0.981   & 0.990   & \textbf{0.986}  & 0.994 & 0.989 & \multicolumn{1}{c}{\textbf{0.992}} \\
\multicolumn{1}{c}{}                                  & Voting  & 0.990    & 0.989  & 0.990    & 0.981   & 0.989  & 0.985  & 0.994 & 0.988 & \multicolumn{1}{c}{0.991} \\
\multicolumn{1}{c}{}                                  & ANN    & 0.991   & 0.986   & 0.989   & 0.979   & 0.990   & 0.984  & 0.990  & 0.989 & \multicolumn{1}{c}{0.990}  \\ 
\hline
\end{tabular}
\end{table*}
%%%%%%%%%%
\begin{table*}
\centering
\caption{Comparison of F1-scores across various input patterns in four-class classification, including differential aperture magnitudes ($ “w1mag1-w1mag3” $ and $ “w1mag1-w1mag3”, “w2mag1-w2mag3” $) vs. exclusion, analyzing their impact on classification performance.}
\begin{tabular}{|l|c|c|c|c|}
\hline
Input Pattern & F1 (NG) & F1 (ELG) & F1 (Star) & F1 (QSO) \\
\hline
$ r, u-g, g-r, r-i, i-z, z-W1, W1-W2 $ & 0.84 & 0.85 & 0.99 & 0.98 \\
$ r, u-g, g-r, r-i, i-z, z-W1, W1-W2, w1mag1-w1mag3 $ & 0.87 & 0.82 & 0.99 & 0.98 \\
$ r, u-g, g-r, r-i, i-z, z-W1, W1-W2, w1mag1-w1mag3, w2mag1-w2mag3 $ & 0.88 & 0.81 & 0.99 & 0.98 \\
\hline
\end{tabular}
\label{tab:aperture}
\end{table*}

%%%%%%%%%%
\begin{table*}
\centering
	\caption{The performance of all ML methods in the four-class experiment for both Samples I and II with optical and infrared information.}
	\label{tab:four-class}
\begin{tabular}{cccccccccccccc}
\hline 
\multicolumn{2}{c}{}                       & \multicolumn{3}{c}{NG} & \multicolumn{3}{c}{ELG} & \multicolumn{3}{c}{Star} & \multicolumn{3}{c}{QSO} \\
Features                          & Method & P         & R         & F1        & P           & R           & F1          & P      & R      & F1     & P      & R      & F1    \\ \hline
\multirow{5}{*}{Optical}        & KNN     & 0.832     & 0.873     & 0.852       & 0.748       & 0.654       & 0.698  & 0.934  & 0.949  & 0.941  & 0.912  & 0.922  & 0.917 \\
                                  & RF        & 0.842     & 0.864     & 0.853     & 0.742       & 0.680       & 0.710       & 0.941  & 0.946  & 0.944  & 0.909  & 0.928  & 0.918 \\
                                  & XGB    & 0.847     & 0.864     & 0.855     & 0.745       & 0.687       & 0.715       & 0.942  & 0.946  & 0.944  & 0.910  & 0.929  & 0.919 \\
                                  & Voting  & 0.843     & 0.869     & 0.855     & 0.749       & 0.678       & 0.712       & 0.940  & 0.948  & 0.944  & 0.911  & 0.928  & 0.920  \\
                                  & ANN    & 0.845     & 0.861     & 0.853     & 0.757       & 0.652       & 0.700       & 0.954  & 0.926  & 0.940   & 0.859 & 0.946  & 0.900   \\ \hline
\multirow{5}{*}{Optical+IR} & KNN     & 0.862     & 0.891     & 0.876       & 0.827       & 0.783       & 0.804   & 0.995  & 0.984  & 0.990  & 0.978  & 0.991  & 0.985 \\
                                  & RF        & 0.874     & 0.886     & 0.880     & 0.827       & 0.802       & 0.815       & 0.995  & 0.988  & 0.991  & 0.979  & 0.992  & 0.985 \\
                                  & XGB    & 0.874     & 0.885     & 0.879     & 0.826       & 0.804       & 0.815       & 0.994  & 0.989  & 0.991  & 0.980   & 0.991  & 0.985 \\
                                  & Voting  & 0.871     & 0.888     & 0.880     & 0.828       & 0.799       & 0.813       & 0.995  & 0.987  & 0.991  & 0.979   & 0.992  & 0.985 \\
                                  & ANN    & 0.870     & 0.888     & 0.879     & 0.832       & 0.785       & 0.808       & 0.991  & 0.989  & 0.990   & 0.971  & 0.991  & 0.981 \\ \hline
\end{tabular}
\end{table*}

Based on the evaluation metrics for the four-class experiment in Table~\ref{tab:four-class}, 
the best F1 for NG is 0.88 for both RF and voting.
There is no better classifier than RF and XGB for ELG with the same F1 values of 0.815.
The star class is well identified by all five classifiers, with the F1 value of 0.991 for RF, 
XGB and voting, and 0.99 for KNN and ANN. As far as QSO is concerned, 
all algorithms show good performance, e.g.,
F1 of KNN, RF, XGB, and voting are equally 0.985 and of ANN is 0.981.
Overall, as compared to other classes, stars can be distinguished with higher performance.

\subsection{Misclassification results}
\begin{figure*}
\centering
        \begin{subfigure}[b]{0.25\textwidth}
                \centering
                \includegraphics[width=\linewidth]{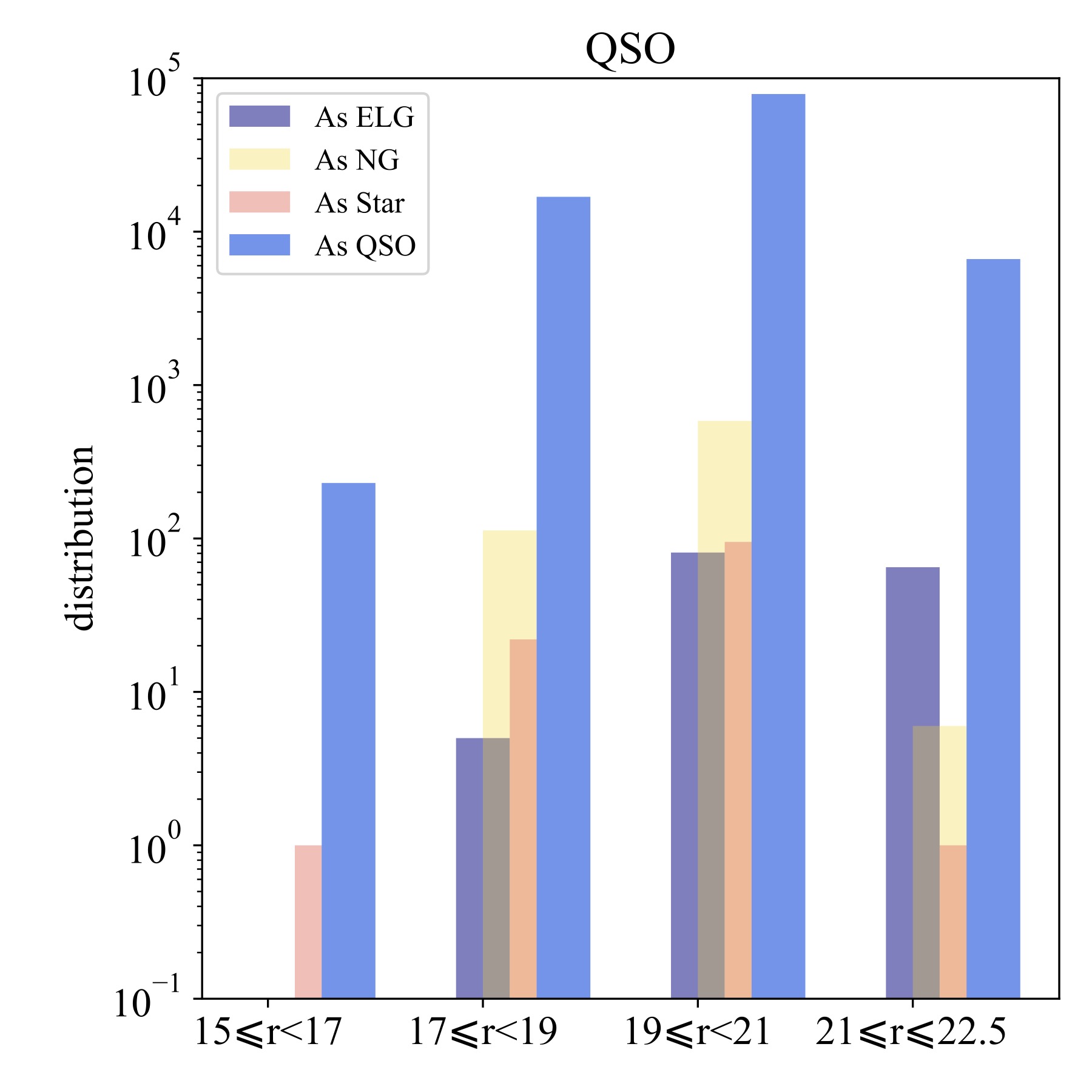}
                \caption{QSO misclassified as other sources.}
                \label{fig:QSO}
        \end{subfigure}\hfill
        \begin{subfigure}[b]{0.25\textwidth}
                \centering
                \includegraphics[width=\linewidth]{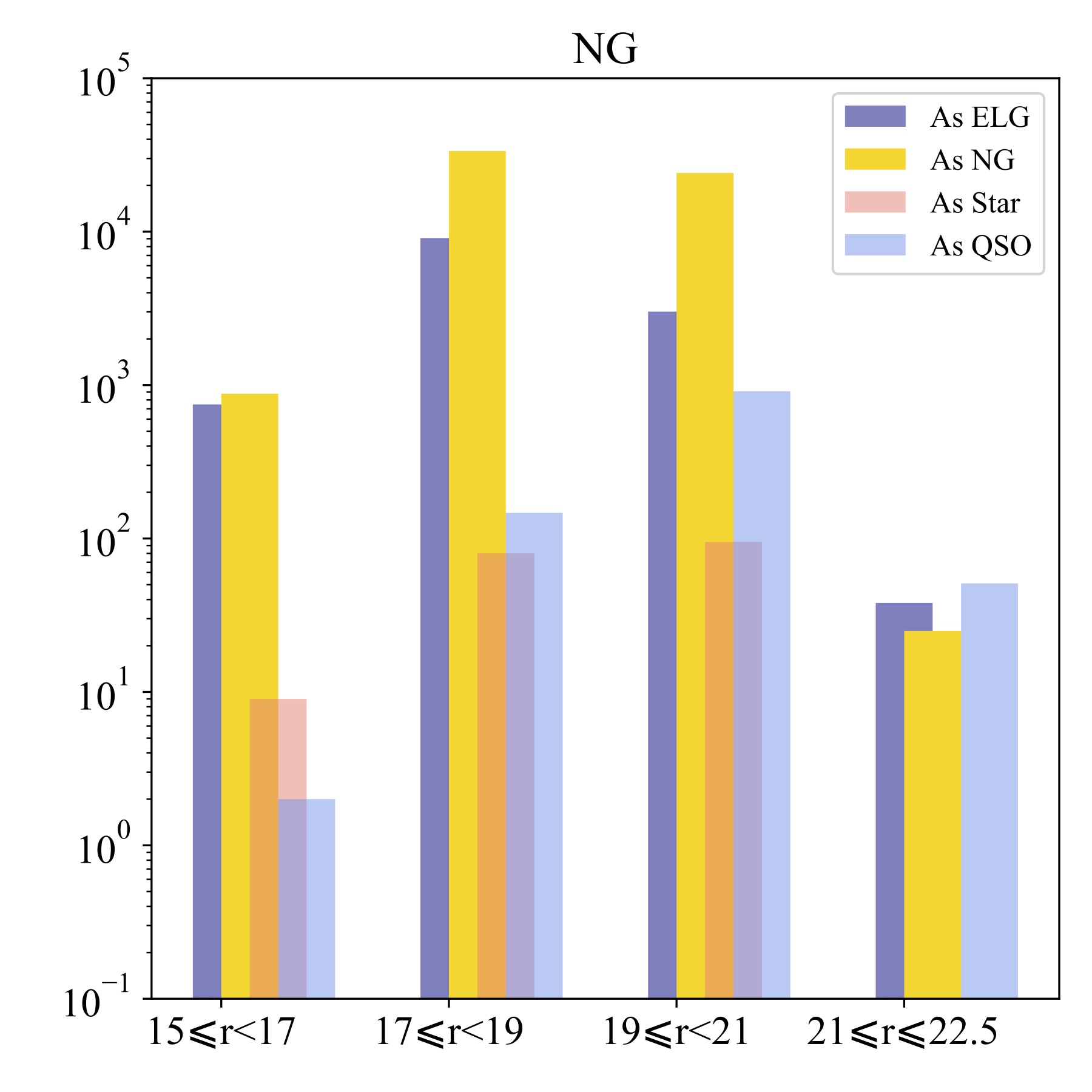}
                \caption{NG misclassified as other sources.}
                \label{fig:NG}
        \end{subfigure}\hfill
        \begin{subfigure}[b]{0.25\textwidth}
                \centering
                \includegraphics[width=\linewidth]{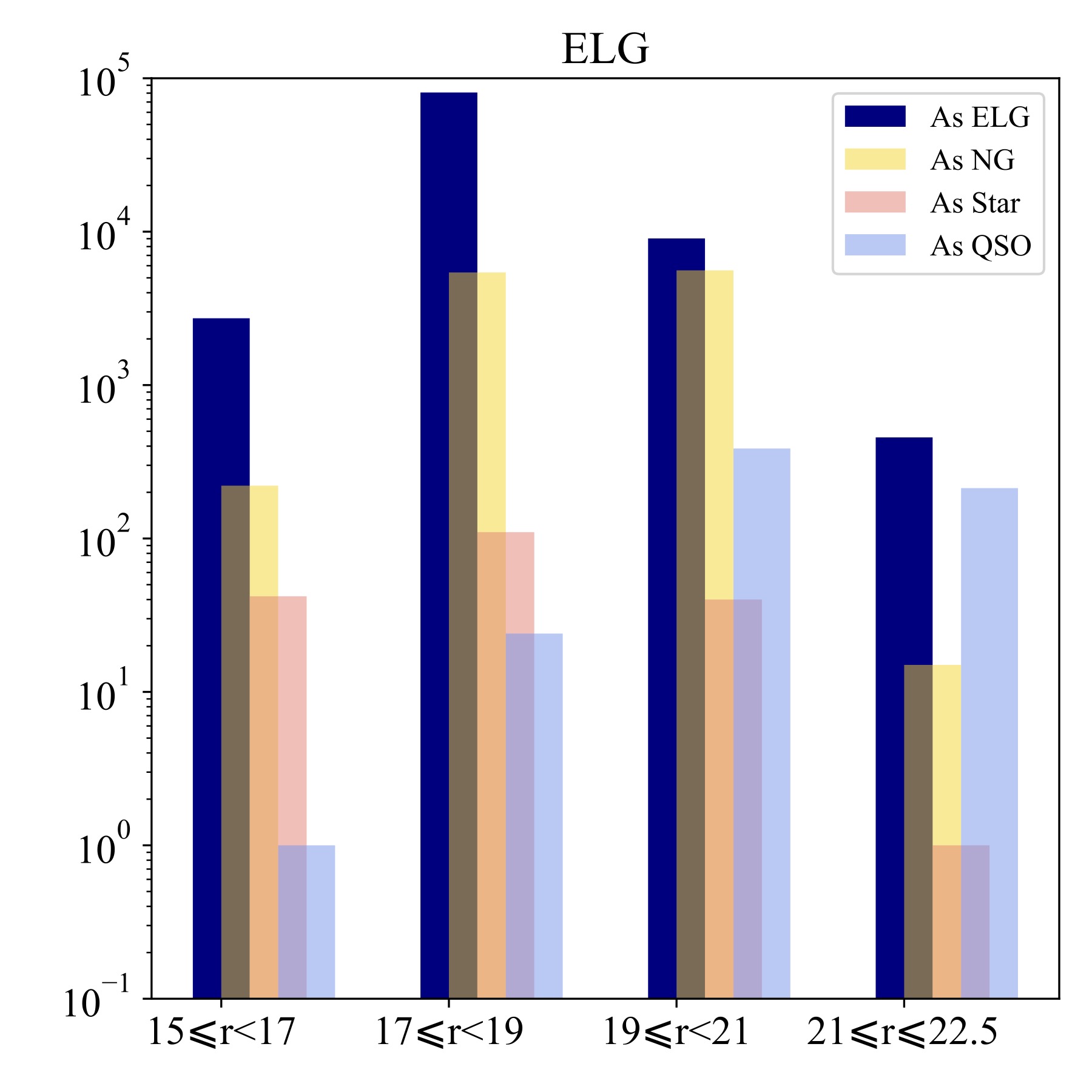}
                \caption{ELG misclassified as other sources.}
                \label{fig:EG}
        \end{subfigure}\hfill
        \begin{subfigure}[b]{0.25\textwidth}
                \centering
                \includegraphics[width=\linewidth]{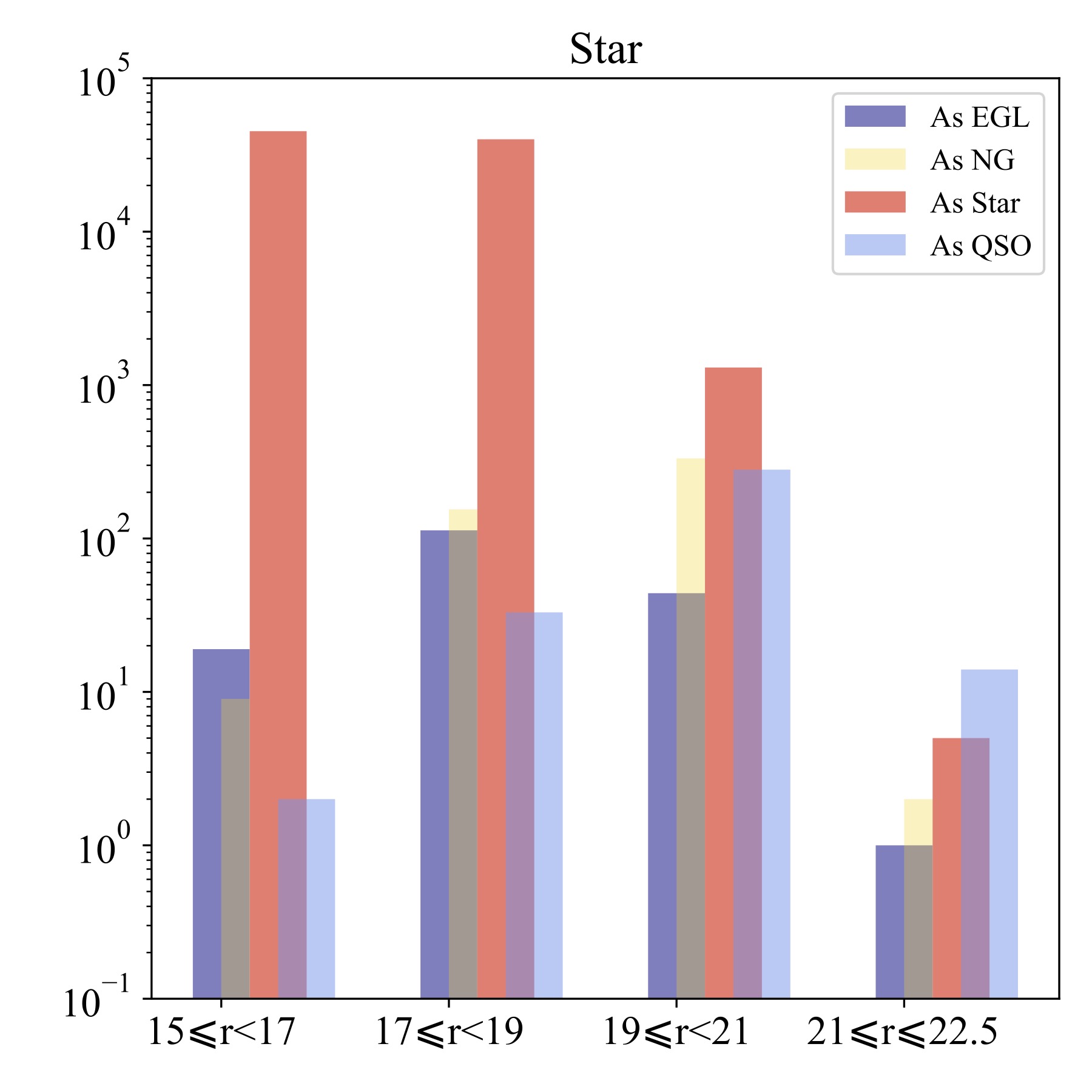}
                \caption{Star misclassified as other sources.}
                \label{fig:star}
        \end{subfigure}
        \caption{Misclassified sources in four bins of $r$ magnitude in the four-class experiment by XGB classifier.}\label{fig:misclass}
\end{figure*}

In order to be able to show a more clear picture of the results across different ranges of $r$ magnitude, 
we divide the sample into four bins of $ 15 \leqslant r < 17 $, $ 17 \leqslant r < 19 $, 
$ 19 \leqslant r < 21 $, and $ 21 \leqslant r \leqslant 22.5 $.
Fig.~\ref{fig:misclass} illustrates the distribution of sources classified correctly and incorrectly in 
the four-class experiment described in Sec.~\ref{subsubsec:second step}.
This plot is based on the results of the XGB classifier.
The QSOs, NGs, ELGs, and stars misclassified as other classes are shown 
in Panels (a), (b), (c), and (d), respectively.
All panels show that sources are classified well in ranges where the number of data is satisfactory.
Accordingly, for all four types of sources, the best 
classification results are obtained in the ranges of $ 17 \leqslant r < 19 $ and $ 19 \leqslant r < 21 $.
Panel~(a) displays good classification of QSOs since there 
are sufficient data in all ranges of $r$ magnitude to allow accurate classification.
As Panel~ (b) shows, NGs are mostly misclassified as ELGs in all bins.
As a result of the scarcity of sources, the classifier performs poorly in $ 21 \leqslant r \leqslant 22.5 $.
In this range, the number of incorrectly classified NGs is higher than that of correctly classified NGs.
According to Panel~(c), ELGs are frequently misclassified as NGs except for 
the range of $ 21 \leqslant r \leqslant 22.5 $ which are mainly classified as QSOs.
NGs and ELGs are challenging to classify (Panels~(b) and (c)); however, they 
can be effectively classified within the ranges of $ 17 \leqslant r < 19 $ and 
$ 19 \leqslant r < 21 $ where there are adequate data.
As shown in Panel~(d), stars are identified well in the bright magnitude ranges, 
$ 17 \leqslant r < 19 $ and $ 19 \leqslant r < 21 $; however, the algorithm 
does not perform well in the faint range of $ 21 \leqslant r \leqslant 22.5 $, 
because there are fewer sources, less than ten, and the classifier has 
a tendency to misclassify them as QSOs, which is consistent with the fact that quasars are generally brighter than stars and galaxies and occupy the majority in the faint magnitude.

\begin{figure}
     \centering
       \begin{minipage}{0.48\textwidth}
     \includegraphics[width=1\linewidth]{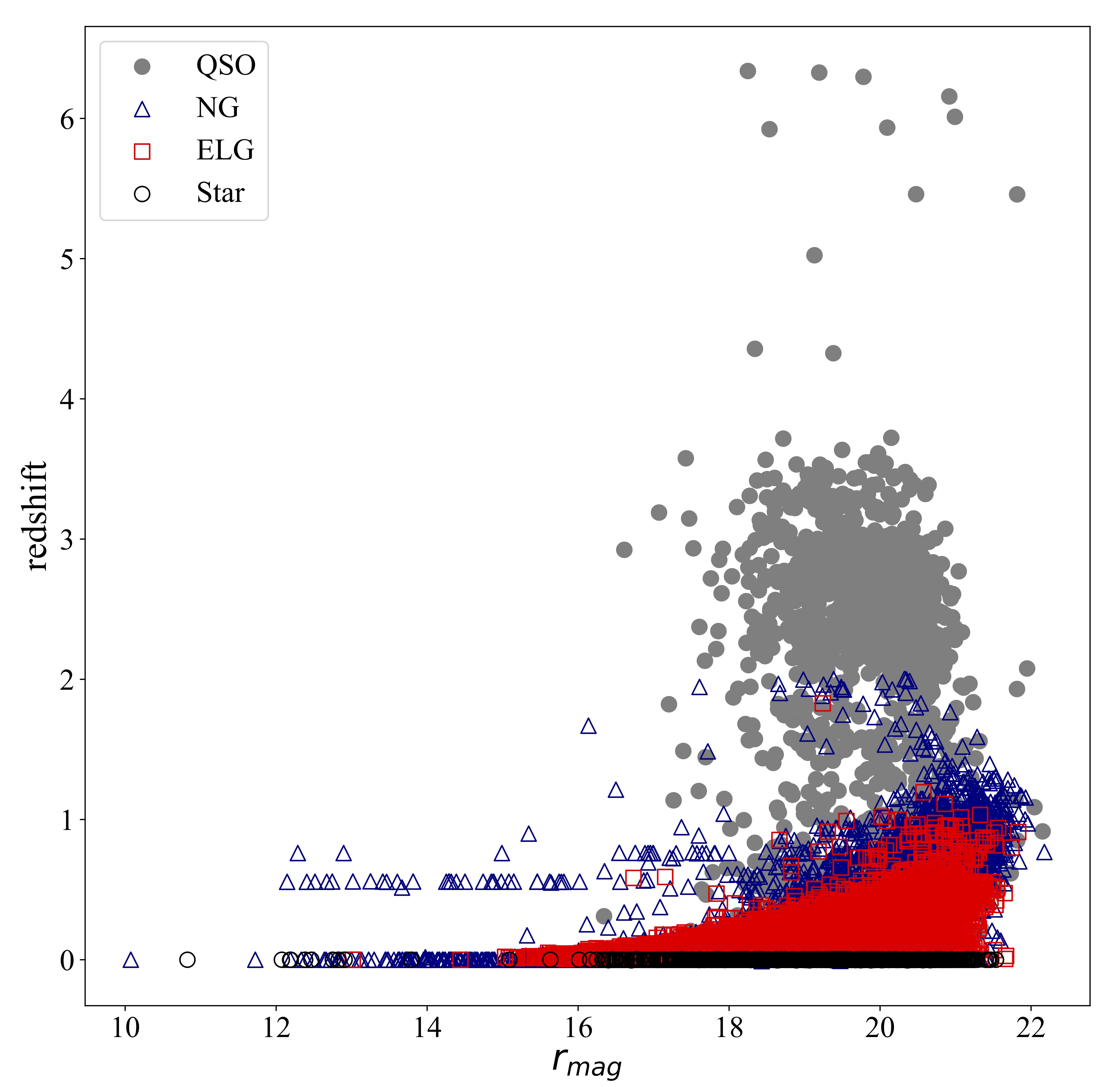}
     \caption{QSOs, NGs, ELGs, and stars misclassified as 
the other three sources are shown with gray circles, blue triangles, 
red squares, and open circles, respectively.}\label{fig:misclassified}
   \end{minipage}\hfill
\end{figure}

We conduct a more detailed analysis of the misclassified sources, focusing specifically on their magnitudes in the $r$ band and redshifts in Fig.~\ref{fig:misclassified}.
QSOs, NGs, ELGs, and stars misclassified as the other three sources are shown with gray circles, blue triangles, red squares, and open circles, respectively.
The impact of classification may be affected by sample imbalance, where the minority class is more likely to be classified as the majority class. For instance, dark/high-redshift sources may be mistakenly classified as bright/low-redshift sources. Fig.~\ref{fig:distribution} highlights an obvious overrepresentation of ELGs over NGs and QSOs at brighter magnitudes, which could potentially influence the classification effect. 
Furthermore, at lower brightness levels and potentially higher redshifts, the ELGs that are incorrectly classified as NGs may be missed due to low signal-to-noise ratios of their emission lines. 

\section{Discussion} \label{sec:discussion}
This study aims to investigate the potential of using photometric parameters 
to identify and classify different types of astronomical objects. 
To evaluate our results, we compare them with other studies that utilize photometric features. 
However, conducting a detailed comparison of results from surveys with varying data selection criteria is not feasible. 
Therefore, we limit our analysis to a general comparison of studies conducted on the SDSS and WISE data that are similar to our work.
Given the differences in the training and validation strategies, we are unable to provide a direct and quantitative comparison of our experimental results with others' works.
Further, there is a paucity of literature that deals with photometry-based 
four-class classification akin to our work, so we have resorted to comparing 
the outcomes of our three-class classification with those of similar studies in the literature.
As an example, \citet{Nakoneczny2021} leveraged SDSS+WISE to train 
machine learning algorithms and generate a QSO catalogue from KIDS data.
According to their findings, infrared information is capable of discriminating 
QSOs efficiently, and using XGB classifier obtained their best average F1-score 
of 96.85 per~cent whereas we obtained an improved average F1-score 
of 98.93 per~cent using the same algorithm (see Table~\ref{tab:three-class}).
As shown in Tables~\ref{tab:RF}-\ref{tab:four-class}, our study is also consistent 
with Nakoneczny and colleagues that infrared information is extremely important for 
classification based on photometric data. Fig.~\ref{fig:feature_importance} 
also turns out that the most important feature is $W1-W2$
and the second important is $z-W1$.
To maximize performance, it is recommended to incorporate both optical and infrared information. 
When both optical and infrared information are used together, the predictability of a source is higher 
compared to using optical information alone. 
In terms of added value, the inclusion of optical colors enhances our classification approach. While WISE data are crucial for successful classification, the incorporation of optical colors improves the overall accuracy and completeness of the results. By combining optical and IR data, we can leverage the complementary nature of these two wavelength regimes, leading to a more comprehensive understanding of the objects' nature and improving classification accuracy. Consequently, classifiers that utilize both 
optical and infrared information tend to be more reliable than those that rely entirely on optical information.

We can also evaluate the effectiveness of our proposed method by comparing it 
with the approach presented by \citet{Clarke2020}, who trained RF classifier 
and obtained an average F1-score of 96.77 per~cent. Despite this impressive performance, 
our approach still outperforms it with an average F1-score of 98.87 per~cent 
in the three-class experiment in both parts one and two of our study (Tables~\ref{tab:RF}-\ref{tab:three-class}).

\citet{Zhang2019} used measurements of optical spectra of galaxies from SDSS for 
the classification of ELGs located at intermediate redshifts into four categories using several 
different ML methods such as KNN, SVM, RF, and MLP.
The AUC of RF algorithm for starforming, composite, AGN,
and low-ionization nuclear emission regions (LINERs) was obtained 98.5 per~cent, 96.5 per~cent, 86.0 per~cent, and 88.4 per~cent, respectively. 
Although only with photometric data not spectroscopic data, the AUC of our RF classifier in the four-class experiment is 97.1 per~cent for ELG ,  96.8 per~cent for NG, 99.4 per~cent for star, and  99.5 per~cent for QSO, respectively.

In comparison to similar studies, our classification approach demonstrates better performance. \citet{Cunha2022}
obtained an average F1 98.13 per~cent using XGBoost, LightGBM, and CatBoost classifiers 
and \citet{Chaini2022} used a combination of photometric information and images from SDSS to 
achieve the best averaged F1-score of 93.3 per~cent through the use of artificial neural networks (ANN) and conventional neural networks (CNN).

Our work also surpasses similar studies that solely focus on infrared features. For instance \citet{Kurcz2016} 
conducted research on the automatic classification of WISE sources into stars, galaxies, and quasars, utilizing support vector machines (SVM). 
By employing the radial Kernel and presenting their findings in Table 2 of their paper, they achieved an average F1-score of 85.7 per~cent.
Overall, our method demonstrates strong reliability and consistent classification accuracy, highlighting its potential for practical applications in relevant fields.

\section{Conclusions} \label{sec:conclusion}

This study examines the application of multiclass classification for the classification of 
astronomical objects with photometric data, and the results are quite promising.
The possibility of identifying reliable astronomical objects exclusively by photometric 
measurements presents an opportunity to select relevant samples for subsequent 
spectroscopic observation and further studies that are adequately characterized. While this study is exploratory, 
it represents a crucial first step towards refining the classifications of different kinds of 
astronomical objects using multiband photometry. To achieve this, we train supervised 
ML classification techniques on spectroscopically classified objects and investigate 
how ML algorithms can provide accurate classifications based on photometric features.
During the course of this study, there are two main parts in the methodology that we adopt.
Initially, we test our algorithm on a balanced dataset. The confidence
of the classification accuracy on a balanced dataset is important to scale the algorithms 
on an unbalanced dataset. This is done when additional experiments are conducted.
For the first part of our concept, we carry out three experiments using RF 
as the method for the classification of astronomical objects in order to be able 
to come up with a reliable process. The purpose of these three experiments 
is to test the ability to classify objects into three, four, and seven different classes.
The composing of three experiments leads us to investigate the possibility of 
categorizing galaxies into normal and emission-line galaxies, and then 
a finer classification of ELGs into AGN, SF, SB, and BL galaxies.
Based upon the seven-class experiment, it is found that a considerable 
number of AGNs have been misclassified as NGs. Therefore, we decide
to merge the AGNs, SFs, SBs, and BLs into a unified class of ELGs. 
Then we perform four-class classification of stars, QSOs, NGs, and 
ELGs in addition to the three-class experiment to separate stars, QSOs, 
and galaxies using a variety of ML methods as the second part of our study.
Using all the existing data, we expand our dataset 
in this part and train the ML models with both optical and 
optical+IR data separately. According to the results, XGB and voting 
perform better when compared to other algorithms.
It is clear from the results that the infrared features are able to substantially 
improve the evaluation metrics in all the experiments. To examine how 
the classification of these sources depends on the brightness 
of the sources, we break up the samples into four bins of magnitude in the $r$ band.
All four classes are generally classified well when there are enough data
in the bins. As far as NGs and ELGs are concerned, 
a majority of faint sources are misclassified as QSOs. In contrast, 
the bright NGs are mainly misclassified as ELGs and vice versa.

Our results indicate that the availability of information, 
brightness of sources, quantity of datasets, sample balance, sample completeness and uncertainty level associated with 
spectroscopic classification of sources in the training set significantly contribute to the 
performance of our methods. Automated algorithms trained on photometric data with 
spectroscopic classification may offer an alternative solution to classical methods 
based on time-consuming spectroscopic observations.
In summary, ML techniques have shown promise in classifying different types of astronomical objects, but their effectiveness can vary depending on various factors such as sample selection, feature selection, types, magnitude choices, and sample size. 
Obviously the performance of a classifier is influenced not only by available information, but also by enough representative training sample. 
With the advent of new surveys, which will provide unprecedented amounts of faint data, robust big data processing will be required to efficiently analyze this information. To meet the demands of these future missions, carefully designed, interpretable, and well-tested ML models will be needed to provide reliable and trustworthy results. The framework presented in the article takes a significant step toward meeting these evolving demands.

Moreover, this study acknowledges the limitations inherent in the datasets used (SDSS and WISE), particularly the relatively poorer image quality of these surveys compared to newer ones like LSST. While our results provide insights into celestial object classification within the context of SDSS and WISE, it is crucial to exercise caution when extrapolating conclusions to surveys with improved image quality. The enhanced sensitivity of newer surveys may present unique opportunities and challenges not addressed here. Researchers using different datasets should carefully consider their specific characteristics and exercise prudence when generalizing findings to surveys with distinct observational properties. Additionally, it is worth noting the potential usefulness of morphological parameters for the LSST and similar surveys, as their quality will be significantly better.

\section*{Acknowledgements}
We are very grateful to the referee's constructive comments and suggestions. This paper is funded by the National Natural Science Foundation of China under grant Nos. 12003021 (F.Z.Z.), 12171385 (L.M.), 12150410308 (A.M.), 12273076 (Y.Z.) and the science research grants from the China Manned Space Project with Nos. CMS-CSST-2021-A04 and CMS-CSST-2021-A06.

We acknowledge the use of SDSS  and WISE databases.
Funding for the Sloan Digital Sky Survey IV has been provided by the Alfred P. Sloan Foundation, the U.S. Department of Energy Office of Science, and the Participating Institutions. SDSS-IV acknowledges
support and resources from the Center for High-Performance Computing at
the University of Utah. The SDSS web site is www.sdss.org.
Funding for the SDSS has been provided by the Alfred P.
Sloan Foundation, the Participating Institutions, the National
Science Foundation, the U.S. Department of Energy, the National Aeronautics and Space Administration, the Japanese
Monbukagakusho, the Max Planck Society, and the Higher
Education Funding Council for England. The SDSS Web
Site is http://www.sdss.org/.

The SDSS is managed by the Astrophysical Research Consortium for the Participating Institutions. The Participating
Institutions are the American Museum of Natural History,
Astrophysical Institute Potsdam, University of Basel, University of Cambridge, Case Western Reserve University, University of Chicago, Drexel University, Fermilab, the Institute for
Advanced Study, the Japan Participation Group, Johns Hopkins University, the Joint Institute for Nuclear Astrophysics,
the Kavli Institute for Particle Astrophysics and Cosmology, the Korean Scientist Group, the Chinese Academy
of Sciences (LAMOST), Los Alamos National Laboratory,
the Max-Planck-Institute for Astronomy (MPIA), the MaxPlanck-Institute for Astrophysics (MPA), New Mexico State
University, Ohio State University, University of Pittsburgh,
University of Portsmouth, Princeton University, the United
States Naval Observatory, and the University of Washington.

This publication makes use of data products from the Wide-field Infrared Survey Explorer, which is a joint project of the University of California, Los Angeles, and the Jet Propulsion Laboratory/California Institute of Technology, funded by the National Aeronautics and Space Administration.

%%%%%%%%%%%%%%%%%%%%%%%%%%%%%%%%%%%%%%%%%%%%%%%%%%
\section*{Data Availability}
The catalogues are adopted from the publicly released SDSS database and ALLWISE database.



%\bibliographystyle{mnras}


%\bsp	% typesetting comment
%\label{lastpage}
\end{document}